\documentclass[12pt]{article}

\usepackage[english]{babel}
\usepackage[utf8]{inputenc}
\usepackage{amsmath}
\usepackage{graphicx}
\usepackage[colorinlistoftodos]{todonotes}
\usepackage{amssymb}
\usepackage{mathtools}
\usepackage{amsthm}
\usepackage{natbib}
\usepackage{tikz}
\usepackage{bbm}
\usepackage{adjustbox}
\usepackage{etoolbox}
\usepackage{enumerate}
\usepackage{pdflscape}
\usepackage{adjustbox}
\usepackage{scrextend}
\usepackage{titling}
\usepackage{setspace}
\usepackage{bm}
\usepackage{mathrsfs}
\usepackage[margin=1.25in]{geometry}
\usepackage[toc,page]{appendix}
\usepackage{rotfloat}
\usepackage[normalem]{ulem}

\numberwithin{equation}{section}

\newcommand{\EE}{\mathbb{E}}
\newcommand{\PP}{\mathbb{P}}

\newcommand{\Cor}{\mathbb{C}\mbox{or}}

\newcommand{\Z}{\mathbb{Z}}
\newcommand{\R}{\mathbb{R}}
\newcommand{\N}{\mathbb{N}}
\newcommand{\iid}{\mbox{\textit{i.i.d.}}}

\providecommand{\X}{\mathbf{X}}

\theoremstyle{definition}
\newtheorem{assumption}{Assumption}
\newtheorem{assumption*}{Assumption}

\theoremstyle{plain}
\newtheorem{theorem}{Theorem}

\theoremstyle{remark}

\author{Alexander Heinemann, Eric Beutner and Stephan Smeekes}

\date{\today}

\title{A General Framework for Prediction in Time Series Models}

\begin{document}

\singlespacing

\begin{titlepage}
\begin{center}

\scshape
\LARGE{\textbf{A General Framework for Prediction in Time Series Models}}
\normalfont
\end{center}
\begin{center}
\par\vspace{1.5cm}
\text{Eric Beutner$^\dagger$}
\hspace{1.5cm}
\text{Alexander Heinemann$^\dagger$}
\hspace{1.4cm}
\text{Stephan Smeekes$^\dagger$}
\par\vspace{2cm}
\text{$^\dagger$Department of Quantitative Economics}
\par\vspace{0.3cm}
\text{Maastricht University}
\par\vspace{0.3cm}
\text{\today}
\vspace{1cm}
\end{center}

\begin{abstract}
In this paper we propose a general framework to analyze prediction in time series models and show how a wide class of popular time series models satisfies this framework. We postulate a set of high-level assumptions, and formally verify these assumptions for the aforementioned time series models. Our framework coincides with that of \cite{beutner2019justification} who establish the validity of conditional confidence intervals for predictions made in this framework. The current paper therefore complements the results in \cite{beutner2019justification} by providing practically relevant applications of their theory.
\end{abstract}

\end{titlepage}

\doublespacing

\section{Introduction}

In time series prediction one is frequently interested in objects that do not only depend on parameters but also on the time series' past. Popular examples are conditional means or conditional variances. Analyzing predictions in this context involves a fundamental issue that is well-recognized in the econometric literature. It stems from the fact that on the one hand one must \textit{condition} on the sample as the past informs about the present and future, yet on the other hand one must treat the data up to now as random to take into account parameter uncertainty. Nevertheless the issue is often ignored in standard practice or bypassed by assuming two independent processes with the same stochastic structure, using one for the conditioning and one for the estimation of the parameters. While the latter is a  mathematically convenient assumption, it is rarely satisfied in practice. An alternative, more realistic approach is based on sample-splitting, in which one splits the sample into two (asymptotically) independent subsamples. 

In this paper we provide a general framework to analyze prediction in time series models. We postulate a set of high-level assumptions under which  \cite{beutner2019justification} (henceforth BHS) establish the validity of conditional confidence intervals for predictions while demonstrating an asymptotic equivalence of two-independent processes and the sample-split approach. We show how a wide class of popular time series models satisfies this framework.  In particular, we consider autoregressive moving-average (ARMA) and generalized autoregressive conditional heteroskedasticity (GARCH) type models and formally verify the postulated high-level assumptions. Therefore the current paper complements the results in BHS by providing practically relevant applicants to their theory. 

The rest of this paper is organized as follows. The general framework to analyze prediction in time series models is proposed in Section \ref{sec:3.0} and an accompanying set of high-level assumption is postulated. In Sections \ref{sec:3.1} and \ref{sec:3.2} we revisit the leading examples of BHS, i.e.\ the simple case of a conditional mean in an AR($1$) and the conditional variance in a GARCH($1,1$) model. In Section \ref{sec:3.4} we focus on the conditional mean in a slightly more general model: the ARMA($1,1$) with drift.  Section \ref{sec:3.5} studies the conditional volatility in a threshold GARCH (T-GARCH) model. Concluding remarks are presented in Section \ref{sec:3.6}.


\section{General Framework}
\label{sec:3.0}

Let $\{X_t\}$ be a univariate stochastic process defined on some probability space $(\Omega,\mathcal{F},\PP)$ and denote the relevant parameter (vector) by $\theta_0$, which belongs to some set $ \Theta \subseteq \mathbb{R}^r$, $r \in \N$. The general framework involves inference on objects, which are a function not only of the parameter but also the the time series' past. Mathematically, such object can be written as follows:
%
\begin{equation} \label{eq:pr_func}
\psi_{T+1} :=
\psi (X_T, X_{T-1}, \ldots ; \theta_0)
\end{equation}
for some function $\psi: \mathbb{R}^\infty \times \Theta \rightarrow \mathbb{R}$. Such prediction function can generally not be determined completely given a sample $X_1,\dots,X_T$. Replacing the unknown presample values by arbitrary starting values $\{s_t\}$, yields the following approximation:
\begin{equation} \label{eq:psi_s}
\psi_{T+1}^s (\X_{1:T}; \theta_0) := \psi (X_T, X_{T-1}, \ldots, X_1, s_0, s_{-1}, \ldots; \theta_0),
\end{equation}
where $\X_{t_1:t_2} = (X_{t_1}, \ldots, X_{t_2})^\prime$ for any integers $1 \leq t_1 \leq t_2 \leq T$. To estimate the prediction function in practice, the standard approach is to replace the unknown parameter $\theta_0$ by an estimator $\hat{\theta}(\X_{1:T})$. Conditioning on the entire sample for the evaluation of the prediction function entails that there is no randomness to account for parameter uncertainty, which highlights the severity of the fundamental issue at hand. The issue is frequently bypassed by making the unrealistic assumption of observing two independent processes, where one is used for the evaluation of the prediction function and the other for parameter estimation. 
%

%
An alternative, more realistic approach is based on splitting the sample into two 
(asymptotically) independent subsamples. The successive decline of the influence of past observations, which motivated the approximation in \eqref{eq:psi_s}, entails that
%
%
\begin{equation} \label{eq:18723876}
\psi_{T+1}^s (\X_{t_1:T}^c; \theta_0) = \psi (X_T, X_{T-1}\ldots, X_{t_1},c_{t_1-1},\ldots,c_{1}, s_0, s_{-1}, \ldots; \theta_0)
\end{equation}
%
%
serves as an approximation for \eqref{eq:psi_s} (and hence for  \eqref{eq:pr_func}) for an appropriate choice of $t_1$. Here $\X_{t_1:T}^{c} = (c_1, \ldots, c_{t_1-1}, X_{t_1}, \ldots, X_{T})^\prime$ is a vector where a subsample is substituted by a sequence of constants $\{c_t\}$, in a similar way as done for the starting values. Denoting the appropriate choice of $t_1$ by $T_P$, which indicates the starting point of the prediction sample, the sample-split estimator is obtained  by replacing $\theta_0$ in \eqref{eq:18723876} by an estimator $\hat{\theta}(\X_{1:T_E})$, where $T_E$ stands for the for the end of the estimation sample. Choosing $T_E$ to  satisfy $1< T_E < T_P \leq T$ yields an estimation subsample that does not overlap with the subsample used for prediction.

Next, we postulate a set of high-level assumptions
 under which  BHS establish the validity of conditional confidence intervals for predictions while demonstrating an asymptotic equivalence of two-independent processes and the sample-split approach.
\begin{assumption} (\textit{General Assumptions})
\label{as:2.0}
\begin{enumerate}[\ref{as:2.0}.a]
\item (\textit{Estimator})  $m_T \big(\hat{\theta}(\X_{1:T}) - \theta_0\big)\overset{d}{\to} G_{\infty}$  as $T \to \infty$ for some cdf $G_{\infty}: \mathbb{R}^r \rightarrow [0,1]$, where $m_T$ is a sequence of normalizing constants with $m_T\to\infty$;

\item (\textit{Differentiability}) $\psi(\:\cdot\:; \theta)$ is continuous on $\Theta$ and twice differentiable on $\mathring{\Theta}$;

\item (\textit{Gradient}) $\Big|\Big|\frac{\partial \psi (X_T, X_{T-1}, \ldots ;\theta_0)}{\partial \theta}\Big|\Big|=O_{p}(1)$, where $||\cdot||$ denotes the Euclidean norm;

\item (\textit{Hessian}) $\sup_{\theta \in \mathscr{V}(\theta_0)}\Big|\Big|\frac{\partial^2 \psi (X_T, X_{T-1}, \ldots ;\theta)}{\partial \theta \partial \theta'}\Big|\Big|=O_{p}(1)$ for some open neighborhood $\mathscr{V}(\theta_0)$ around $\theta_0$;

\item (\textit{Initial Condition}) Given sequences $\{s_t\}$ and $\{c_t\}$, we have
\begin{align*}
m_T\big(\psi_{T+1}^s (\X_{t_1:T}^c; \theta_0) - \psi (X_T, X_{T-1}, \ldots ;\theta_0)\big)=o_{p}(1),\\
\bigg|\bigg|\frac{\partial \psi_{T+1}^s (\X_{t_1:T}^c; \theta_0)}{\partial \theta} - \frac{\partial \psi (X_T, X_{T-1}, \ldots ;\theta_0)}{\partial \theta}\bigg|\bigg|=o_{p}(1),\\
\sup_{\theta \in \mathscr{V}(\theta_0)}\bigg|\bigg|\frac{\partial^2 \psi_{T+1}^s (\X_{t_1:T}^c; \theta)}{\partial \theta \partial \theta'} - \frac{\partial^2 \psi (X_T, X_{T-1}, \ldots ;\theta)}{\partial \theta \partial \theta'}\bigg|\bigg|=o_{p}(1)
\end{align*}
for any $t_1 \geq 1$ such that $(T-t_1) / l_T \rightarrow \infty$ as $T \to \infty$ and for some model-specific $l_T$ with $l_T \rightarrow \infty$.
\end{enumerate}
\end{assumption}

\begin{assumption} (\textit{Two Independent Processes}) \label{as:2.1}
\quad
\begin{enumerate}[\ref{as:2.1}.a]
\item (\textit{Existence}) $\{Y_t\}$ is a process defined on $(\Omega,\mathcal{F},\PP)$, distributed as $\{X_t\}$;

\item(\textit{Independence}) $\{Y_t\}$ is independent of $\{X_t\}$.
\end{enumerate}
\end{assumption}

\begin{assumption} \textit{(SPL Estimator)}
\label{as:2.2}
\begin{enumerate}[\ref{as:2.2}.a]
\item (\textit{Rates}) The functions $T_P:\mathbb{N} \to \mathbb{N}$ and $T_E:\mathbb{N} \to \mathbb{N}$ satisfy $T_E (T) < T_P(T)$ for all $T$, while $\frac{T - T_P (T)}{l_T} \rightarrow \infty$ and $m_{T_E (T)}/ m_T \rightarrow 1$ as $T \rightarrow \infty$;
\item (\textit{Strict Stationarity}) $\{X_t\}$ is a strictly stationary process;
\item (\textit{Weak Dependence}) $\{X_t\}$ satisfies for each bounded, real-valued Lipschitz function $h$ on $\R^r$
\begin{align*}
\int h \:d \Big(G_{T_E}^{SPL}(\cdot|\mathcal{I}_{T_P:T})- G_{T_E}^{SPL}\Big) \overset{p}{\to}0 \qquad \text{as}\qquad T \to \infty,
\end{align*}
where $G_{T_E}^{SPL}$ denotes the unconditional cumulative distribution function (cdf) of $m_{T_E} \big(\hat{\theta}(\X_{1:T_E}) - \theta_0\big)$ and $G_{T_E}^{SPL}(\cdot|\mathcal{I}_{T_P:T})$ the corresponding conditional cdf given the $\sigma$-algebra $\mathcal{I}_{T_P:T}=\sigma(X_t:T_P\leq t\leq T)$.
\end{enumerate}
\end{assumption}

\begin{assumption}{\textit{(CDF Estimator)}} \label{as:2.4} Let $\widehat{G}_T (\cdot)$ denote a random ($r$-dimensional) cdf as a function of $\X_{1:T}$, used to estimate $G_{\infty}$. Then $\int h \:d \widehat{G}_T (\cdot) \overset{p}{\to}\int h \:d G_{\infty}$ as $T \to \infty$ for each bounded, real-valued Lipschitz function $h$ on $\R^r$.
\end{assumption}

\begin{assumption}{\textit{(Normality)}}
\label{as:2.5}
Let $G_{\infty}$ be the cdf of the $N(0,\Upsilon_0)$ distribution with $\Upsilon_0=\Upsilon(\theta_0,\xi_0)$ and assume there exist $\hat{\Upsilon}(\X_{1:T})$ 
 converging in probability to $\Upsilon_0$.
\end{assumption}
Assumption \ref{as:2.0} ensures that the prediction function is well behaved and that one can estimate the parameter it depends on. Whereas Assumption \ref{as:2.1} formalizes the unrealistic two-independent-processes assumption, the stationarity and weak dependence condition in Assumption \ref{as:2.2} allow to split the sample into (asymptotically) independent and identical subsamples. The consistent estimation of the asymptotic distribution of the parameter estimator, $G_\infty$, is stated in Assumption \ref{as:2.4}, which simplifies in the case of asymptotic normality (Assumption \ref{as:2.5}).

In the following sections we formally verify the high-level assumptions stated above for a wide class of popular time series models satisfying this framework. Since the subsequently considered ARMA and GARCH models exhibit an exponential decay in memory we henceforth set $l_T = \log T$. Further, we constrain ourselves to $\sqrt{T}$-consistent estimators of the parameters such that $m_T=\sqrt{T}$ throughout the paper.

\section{Conditional Mean in an AR(1)}
\label{sec:3.1}

\subsection{Model Description}
\label{sec:3.1.1}

An autoregressive model represents a process in terms of its lagged value(s) and some stochastic innovation process. The first order autoregressive process without drift is defined by the following recursion
\begin{align}
\label{eq:3.1.1}
X_t =  \beta_0 X_{t-1}+\varepsilon_t\:,
\end{align}
for $t \in \Z$, where the parameter $\beta_0 \in \Theta$ satisfies $|\beta_0|<1$  and $\{\varepsilon_t\}$ is a sequence of innovations. Subsequently, we make the following assumptions.
\begin{assumption}{\textit{(AR$(1)$-Model)}}
\label{as:3.1}
\begin{enumerate}
\item[\ref{as:3.1}.1] \textit{(Compactness)} $\Theta$ is compact;
\item[\ref{as:3.1}.2] \textit{(Interior)} $\beta_0 \in \mathring{\Theta}$, where $\mathring{\Theta}$ denotes the interior of $\Theta$;
\item[\ref{as:3.1}.3] \textit{(Causality)} $|\beta|< 1$ for all $\beta \in \Theta$;
\item[\ref{as:3.1}.4] \textit{(Innovations)}
$\varepsilon_t$ are $\iid$ from an absolutely continuous distribution with respect to the Lebesgue measure on $\R$ satisfying  $\EE[\varepsilon_t]=0$,  $\EE[\varepsilon_t^4]<\infty$ and having a Lebesgue density strictly positive on $\R$;
\end{enumerate}
\end{assumption}

$\Theta$ is assumed to be compact in Assumption \ref{as:3.1}.1, which holds true, for instance, if it is of the form $\Theta=\big\{\beta'\in \R: |\beta|\leq 1-\delta\big\}$, where $\delta>0$ is a sufficiently small constant. Assumption \ref{as:3.1}.2 states that the true parameter vector lies in the interior of the parameter set and is necessary to obtain asymptotic normality of the parameter estimator. The causality condition is stated in \ref{as:3.1}.3. Assumption \ref{as:3.1}.4 imposes further restrictions on the distribution of the innovation process. 
Next, we turn to the estimation of the model.

\subsection{Estimation}
\label{sec:3.1.2}

To estimate the model in equation \eqref{eq:3.1.1}, we employ the OLS estimator given by
\begin{align}
\label{eq:3.1.2}
\hat{\beta}(\mathbf{X}_{1:T})= \sum_{t=2}^T X_{t}X_{t-1}/ \sum_{t=2}^T X_{t-1}^2
\end{align}
As the sample size grows large, the OLS estimator approaches a normal distribution under regulatory conditions.

\begin{theorem}{\citep{hamilton1994time}} \label{thm:3.1}
Under Assumption \ref{as:3.1}
\begin{align}
\label{eq:3.1.3}
\sqrt{T}\big(\hat{\beta}(\mathbf{X}_{1:T})-\beta_0\big) \overset{d}{\to}N(0, \sigma_{\beta}^2)
\end{align}
with $\sigma_{\beta}^2=1-\beta_0^2$.
\end{theorem}

\subsection{Mapping}
\label{sec:3.1.3}

The mapping of the AR(1) process into the general framework is straightforward: $\beta_0$ corresponds to $\theta_0$ and the conditional mean of $X_{T+1}$ is equal to
\begin{align}
\label{eq:3.1.4}
\begin{split}
\psi_{T+1}=\psi(X_T,X_{T-1},\dots;\theta_0)=\beta_0 X_T.
\end{split}
\end{align}
%


\subsection{Verification of Assumptions}
\label{sec:3.1.4}

\subsubsection*{Assumption \ref{as:2.0}}

For Assumption \ref{as:2.0}.a to be met, we consider the OLS estimator in \eqref{eq:3.1.2}, whose asymptotic distribution is specified in Theorem \ref{thm:3.1}.\\

\noindent
 As the function $\psi(\dots;\theta)$ given in \eqref{eq:3.1.4} is continuous on $\Theta$ and twice differentiable on $\mathring{\Theta}$, Assumption \ref{as:2.0}.b is met.\\

\noindent
Consider Assumption \ref{as:2.0}.c and notice that the gradient simplifies to
\begin{align*}
\frac{\partial \psi(X_T,X_{T-1},\dots;\theta_0)}{\partial \theta}=X_T.
\end{align*}
Clearly, $X_T$ is $O_p(1)$  since the process $\{X_t\}$ is strictly stationary; see also Assumption \ref{as:2.2}.c , which is verified below.\\

\noindent
The condition in Assumption \ref{as:2.0}.d is met as 
\begin{align*}
\sup_{\theta \in \mathscr{V}(\theta_0)}\bigg|\bigg|\frac{\partial^2 \psi(X_T,X_{T-1},\dots;\theta)}{\partial \theta \partial \theta'}\bigg|\bigg| = 0.\\
\end{align*}

\noindent
Regarding Assumption \ref{as:2.0}.e, we obtain for $t_1<T$
\begin{align*}
m_T\Big(\psi_{T+1}^s (\X_{t_1:T}^c; \theta_0) - \psi (X_T, X_{T-1}, \ldots ;\theta_0)\Big)=\sqrt{T}\big(\beta_0 X_T-\beta_0 X_T\big)=0
\end{align*}
and
\begin{align*}
\bigg|\bigg|\frac{\partial \psi_{T+1}^s (\X_{t_1:T}^c; \theta_0)}{\partial \theta} - \frac{\partial \psi (X_T, X_{T-1}, \ldots ;\theta_0)}{\partial \theta}\bigg|\bigg|= |X_T-X_T|=0
\end{align*}
as well as
\begin{align*}
\sup_{\theta \in \mathscr{V}(\theta_0)}\bigg|\bigg|\frac{\partial^2 \psi_{T+1}^s (\X_{t_1:T}^c; \theta)}{\partial \theta \partial \theta'} - \frac{\partial^2 \psi (X_T, X_{T-1}, \ldots ;\theta)}{\partial \theta \partial \theta'}\bigg|\bigg|=|0-0|=0,
\end{align*}
which completes the verification of Assumption \ref{as:2.0}.



\subsubsection*{Assumption \ref{as:2.2}}

The condition in Assumption \ref{as:2.2}.a is satisfied for instance by $T_E(T)\sim T-\lfloor T^b \rfloor$ and  $T_P(T)\sim T-\lfloor T^a \rfloor$ with $0<a<b<1$, where $\lfloor x \rfloor$ denotes the largest integer not exceeding $x$.\\

\noindent
The process $\{X_t\}$ is strictly stationary since $|\beta_0|<1$ and $\EE\log^+|\varepsilon_t|\leq \EE|\varepsilon_t|<\infty$, where $\log^+x = \max\{\log x,0\}$  (\citeauthor{bougerol1992stationarity}, \citeyear{bougerol1992stationarity}, Thm.\ 4.1).\\

\noindent
The process $\{X_t\}$ is $\beta$-mixing with exponential decay  (\citeauthor{mokkadem1988mixing}, \citeyear{mokkadem1988mixing}, Thm.\ 1'). As $\beta$-mixing implies $\alpha$-mixing (cf.\ \citeauthor{bradley2005basic}, \citeyear{bradley2005basic}), Assumption \ref{as:2.2}.c  is met with regard to remark 3 of BHS and noting that $T_P(T)-T_E(T)\sim \lfloor T^b \rfloor - \lfloor T^a \rfloor \to \infty$ as $T \to \infty$. For alternative mixing results we refer to \citeauthor{davidson1994stochastic} (\citeyear{davidson1994stochastic}, Thm.\ 14.9) or \citeauthor{andrews1983first} (\citeyear{andrews1983first}, Thm.\ 1).\\

\subsubsection*{Assumptions \ref{as:2.4} and \ref{as:2.5}}

Assumption \ref{as:2.4} is implied by Assumption \ref{as:2.5}, which, in turn,  is verified by Theorem \ref{thm:3.1} and $\hat{\sigma}_\beta^2(\mathbf{X}_{1:T})=1-\hat{\beta}(\mathbf{X}_{1:T})^2\overset{p}{\to}\sigma_\beta^2$.

\subsubsection*{Assumptions within Corollary 1 of BHS}

We show  $1/\hat{\upsilon}_T^{2IP}=O_p(1)$. By independence of $\{\varepsilon_t\}_{t \in \Z}$, the law of $X_T=\sum_{k=0}^\infty\beta_0^k \varepsilon_{T-k}$ is equal to $\mathscr{L}(X_T)=\mathscr{L}(\varepsilon_T)*\mathscr{L}(\beta_0\varepsilon_{T-1})*\mathscr{L}(\beta_0^2\varepsilon_{T-2})*\dots$ As $\mathscr{L}(\varepsilon_t)$ is continuous and non-degenerate, so is $\mathscr{L}(X_T)$, which does not dependent of $T$ as $\{X_t\}_{t \in \Z}$ is strictly stationary. It follows that $X_T$ is bounded away from zero. Further, write $\hat{\upsilon}_T^{2IP}=X_T^2\hat{\sigma}_\beta^2(\mathbf{X}_{1:T})= X_T^2 \sigma_\beta^2+S_T$ and note that $S_T=X_T^2\big(\hat{\sigma}_\beta^2(\mathbf{X}_{1:T})-\sigma_\beta^2\big)=o_p(1)$. For every $\epsilon>0$, we have
\begin{align*}
\PP\big[\hat{\upsilon}_T^{2IP}\geq\epsilon\big]\geq &\PP\Big[X_T^2 \sigma_\beta+S_T\geq\epsilon \cap |S_T|\leq \epsilon\Big]\\
\geq &\PP\Big[X_T^2 \sigma_\beta^2\geq 2\epsilon \cap |S_T|\leq \epsilon\Big]\\
\geq &\PP\big[X_T^2 \sigma_\beta^2\geq 2\epsilon\big] -\PP\big[|S_T|> \epsilon\big],
\end{align*}
where the last inequality follows from $\PP[A\cap B]\geq \PP[A]-\PP[B^c]$. Fix $\delta>0$; since $X_T$ and hence $X_T^2$ are bounded away from zero and $\sigma_\beta^2>0$, there exists an $\epsilon=\epsilon(\delta)$ such that $\PP\big[X_T^2 \geq 2\epsilon/\sigma_\beta^2\big]\geq 1-\delta/2$. For such $\epsilon$, there exists an $\bar{T}=\bar{T}\big(\epsilon(\delta),\delta\big)=\bar{T}(\delta)$ such that $\PP\big[|S_T|> \epsilon\big]<\delta/2$ for all $T\geq \bar{T}$ since $S_T=o_p(1)$. It follows that $\PP\big[\hat{\upsilon}_T^{2IP}\geq\epsilon\big]\geq 1-\delta$ for all $T\geq \bar{T}$. As $\delta>0$ was arbitrarily chosen, this completes the proof of  $\hat{\upsilon}_T^{2IP}$ being bounded away from zero. The proof of $\hat{\upsilon}_T^{SPL}$ being bounded away from zero is analogous and hence omitted.

\section{Conditional Variance in a GARCH(1,1)}
\label{sec:3.2}

\subsection{Model Description}
\label{sec:3.2.1}

Autoregressive conditional heteroscedasticity  models were originally introduced by \cite{engle1982autoregressive} and extended to GARCH models by \cite{bollerslev1986generalized}. The model reflects the predominant characteristics of financial returns justifying its popularity among practitioners. The model's temporal dependence structure captures the slow decaying autocorrelations of absolute financial returns,  also known as volatility clustering. The GARCH$(1,1)$ process $\{X_t\}$ is defined by 
\begin{align}
\label{eq:3.2.1}
\begin{split}
X_t =& \:\sigma_t \varepsilon_t\\
\sigma_t^2 =& \: \omega_0 + \alpha_0 X_{t-1}^2+ \beta_0 \sigma_{t-1}^2
\end{split}
\end{align}
for all $t \in \Z$, where $\theta_0 =(\omega_0,\alpha_0,\beta_0)'$ are non-negative parameters in a parameter set $\Theta$ and $\{\varepsilon_t\}$ is a sequence of innovations. In the traditional GARCH model, \cite{bollerslev1986generalized} assumed the innovations $\{\varepsilon_t\}$ to be independent following a standard normal distribution. The normality assumption is commonly relaxed to account for stylized statistical properties of financial returns such as \textit{skewness} due to leverage effects and \textit{kurtosis}, also known as fat tails. We denote by $\theta=(\omega,\alpha,\beta)'$ a generic  parameter vector and subsequently make the following assumptions:
\begin{assumption}{\textit{(GARCH(1,1)-Model)}}
\label{as:3.2}
\begin{enumerate}
\item[\ref{as:3.2}.1] \textit{(Compactness)} $\Theta$ is compact;
\item[\ref{as:3.2}.2] \textit{(Interior)} $\theta_0$ belongs to $\mathring{\Theta}$;
\item[\ref{as:3.2}.3] \textit{(Non-negativity)}  $\omega>0$, $\alpha\geq 0$  and  $ \beta \geq 0$ for all $\theta \in \Theta$;
\item[\ref{as:3.2}.4] \textit{(Strict Stationarity)} $\EE\big[\ln(\alpha_0\varepsilon_t^2+\beta_0)\big]<0$ and $\beta<1$ for all $\theta \in \Theta$;
\item[\ref{as:3.2}.5] \textit{(Roots)} $\alpha_0 z>0$ and $1-\beta_0 z>0$ have no common root, and $\alpha_0>0$;
\item[\ref{as:3.2}.6] \textit{(Innovations)} $\varepsilon_t$ are $\iid$ from an absolutely continuous distribution with respect to the Lebesgue measure on $\R$ satisfying  $\EE[\varepsilon_t]=0$,  $\EE[\varepsilon_t^2]=1$ and $\EE[\varepsilon_t^4]<\infty$ and having a Lebesgue density strictly positive in a neighborhood of zero;
\end{enumerate}
\end{assumption}
$\Theta$ is assumed to be compact in Assumption \ref{as:3.2}.1, which holds true, for instance, if it is of the form $\Theta=[\delta,1/\delta]\times[0,1/\delta]\times[0,1-\delta]$, where $\delta \in (0,1)$ is a sufficiently small constant. Assumption \ref{as:3.2}.2 states that the true parameter vector lies in the interior of the parameter set and is necessary to obtain asymptotic normality of the parameter estimator.  The non-negativity constraints in \ref{as:3.2}.3 are standard ensuring the conditional variance to be strictly positive. Assumption  \ref{as:3.2}.4 is necessary and sufficient for $\{X_t\}$ being strictly stationary (cf.\ \citeauthor{francq2011garchbook}, \citeyear{francq2011garchbook}, Thm.\ 2.1). The root condition in \ref{as:3.2}.5 guarantees that the GARCH model is irreducible. Assumption \ref{as:3.2}.6 imposes further restrictions on the moments and density of the innovation process. 
Next, we turn to the estimation of the model in \eqref{eq:3.2.1}.

\subsection{Estimation}
\label{sec:3.2.2}

We consider the quasi maximum likelihood (QML) estimator proposed by \cite{francq2004maximum} to estimate the GARCH($1,1$) model. For a generic $\theta \in \Theta$ we set
\begin{align}
\label{eq:877590858}
 \sigma_{t+1}^2(\theta) = \sum_{k=0}^\infty \beta^k  \big(\omega + \alpha X_{t-k}^2\big)
\end{align}
and note that $\sigma_{t+1}^2=\sigma_{t+1}^2(\theta_0)$. Replacing the unknown presample observations by arbitrary values, say $s_t$, $t\leq 0$, we denote the modified version of \eqref{eq:877590858} by $\tilde{\sigma}_{t+1}^2(\theta)$.
%
%
Then the QML estimator of $\theta_0$ is defined as any measurable solution $\hat{\theta}(\mathbf{X}_{1:T})$ of
\begin{align}
\label{eq:3.2.4}
\hat{\theta}(\mathbf{X}_{1:T})=&\arg\max_{\theta \in \Theta} \tilde{L}_T(\theta;\mathbf{X}_{1:T})
\end{align}
with
\begin{align*}
\tilde{L}_T(\theta;\mathbf{X}_{1:T})=\prod_{t=1}^T\frac{1}{\sqrt{2\pi \tilde{\sigma}_t^2(\theta)}}\exp\left(-\frac{X_t^2}{2\tilde{\sigma}_t^2(\theta)}\right).
\end{align*}
Assumption \ref{as:3.2} implies that the estimator follows asymptotically a normal distribution.

\begin{theorem}{\citep{francq2004maximum}}
\label{thm:3.2} Under Assumption \ref{as:3.2}
\begin{align}
\label{eq:3.2.5}
\sqrt{T}\big(\hat{\theta}(\mathbf{X}_{1:T})-\theta_0\big) \overset{d}{\to}N\big(0, \Upsilon_0\big),
\end{align}
where $\Upsilon_0 = \big(\EE[\varepsilon_t^4]-1\big)\:\EE\left[\frac{1}{\sigma_t^4}\frac{\partial \sigma_t^2(\theta_0)}{\partial \theta}\frac{\partial \sigma_t^2(\theta_0)}{\partial \theta'}\right]^{-1}$ and
$\sigma_t^2(\theta)$ is given in \eqref{eq:877590858}.
\end{theorem}
It is worth stressing that $\Upsilon_0$ does not only depend on $\theta_0$ but also on some nuisance parameters such as $\EE[\varepsilon_t^4]$.

\subsection{Mapping}
\label{sec:3.2.3}

Having described the model and its estimation, we turn to map the model into the general setup. The conditional variance $\sigma_{T+1}^2$ is equal to
\begin{align}
\label{eq:3.2.7}
\psi_{T+1}=\psi(X_T,X_{T-1},\dots;\theta_0)
=& \sum_{k=0}^\infty \beta_0^k  \big(\omega_0 + \alpha_0 X_{T-k}^2\big).
\end{align}
%
%
%
To verify Assumption \ref{as:2.0} the first and second derivatives of $\psi(X_T,X_{T-1},\dots;\theta)$ w.r.t.\ $\theta$ are needed. The first order derivatives are
\begin{align*}
\frac{\partial \psi(X_T,X_{T-1},\dots;\theta)}{\partial \omega}=& \frac{1}{1-\beta},\\
\frac{\partial \psi(X_T,X_{T-1},\dots;\theta)}{\partial \alpha}=& \sum_{k=0}^{\infty}\beta^k X_{T-k}^2,\\
\frac{\partial \psi(X_T,X_{T-1},\dots;\theta)}{\partial \beta}=& \sum_{k=1}^{\infty}k\beta^{k-1} \big(\omega + \alpha X_{T-k}^2 \big),
\end{align*}
whereas the second order derivatives are given by
\begin{align*}
\frac{\partial^2 \psi(X_T,X_{T-1},\dots;\theta)}{\partial \omega^2}=&\: 0,\\
\frac{\partial^2  \psi(X_T,X_{T-1},\dots;\theta)}{\partial \omega \partial \alpha}=&\: 0,\\
\frac{\partial^2 \psi(X_T,X_{T-1},\dots;\theta)}{\partial \alpha^2}=&\: 0,\\
\frac{\partial^2 \psi(X_T,X_{T-1},\dots;\theta)}{\partial \omega \partial \beta}=&\: -\frac{1}{(1-\beta)^2},\\
\frac{\partial^2 \psi(X_T,X_{T-1},\dots;\theta)}{\partial \alpha \partial \beta}=&\sum_{k=1}^{\infty} k \beta^{k-1} X_{T-k}^2,\\
\frac{\partial^2 \psi(X_T,X_{T-1},\dots;\theta)}{\partial \beta^2}=& \sum_{k=2}^{\infty}k(k-1)\beta^{k-2} \big(\omega+  \alpha X_{T-k}^2 \big).
\end{align*}

\subsection{Verification of Assumptions}
\label{sec:3.2.4}

Before turning to the verification of the high-level assumptions, note that the strict stationarity condition implies the existence of fractional moments: there exists an $s\in (0,1)$ such that $\EE X_t^{2s}<\infty$ (\citeauthor{nelson1990stationarity}, \citeyear{nelson1990stationarity}, Thm.\ 2). For such $s\in(0,1)$ the following elementary inequalities hold: $(a+b)^s\leq a^s+b^s$ for all $a,b\geq 0$ and $c^s\leq c$ for all $c\geq 1$. 

\subsubsection*{Assumption \ref{as:2.0}}

For Assumption \ref{as:2.0}.a to be met, we consider the QML estimator of \cite{francq2004maximum}, whose asymptotic distribution is specified in Theorem \ref{thm:3.2}.\\

\noindent
As the function $\psi(\dots;\theta)$, given in \eqref{eq:3.2.7}, is continuous on $\Theta$ and twice differentiable on $\mathring{\Theta}$, Assumption \ref{as:2.0}.b is satisfied.\\

\noindent
Consider Assumption \ref{as:2.0}.c and note that
\begin{align*}
\frac{\partial \psi(X_T,X_{T-1},\dots;\theta_0)}{\partial \omega}= \frac{1}{1-\beta_0}
\end{align*}
is trivally $O(1)$. For showing $\frac{\partial \psi(X_T,X_{T-1},\dots;\theta_0)}{\partial \alpha}=O_p(1)$, we need to find a finite $M$ for every $\epsilon>0$ such that $\PP\big[\big|\frac{\partial \psi(X_T,X_{T-1},\dots;\theta_0)}{\partial \alpha}\big|\geq M\big]<\epsilon$ for $T$ sufficiently large. Employing the Markov inequality, we obtain
\begin{align*}
&\PP\bigg[\bigg|\frac{\partial \psi(X_T,X_{T-1},\dots;\theta_0)}{\partial \alpha}\bigg|\geq M\bigg]\leq \frac{1}{M^s}\EE \bigg[\bigg(\sum_{k=0}^\infty\beta_0^{k}X_{T-k}^2\bigg)^s\bigg]\\
\leq & \frac{1}{M^s}\sum_{k=0}^\infty \beta_0^{sk}\EE X_{t}^{2s}= \frac{\EE X_t^{2s}}{(1-\beta_0^s)M^s}
\end{align*}
such that $M>\Big(\frac{\EE X_t^{2s}}{(1-\beta_0^s)\epsilon}\Big)^{1/s}$ gives the desired result. Similarly, we get 
\begin{align*}
\begin{split}
&\PP\bigg[\bigg|\frac{\partial \psi(X_T,X_{T-1},\dots;\theta_0)}{\partial \beta}\bigg|\geq M\bigg] \leq  \frac{1}{M^s}\EE \bigg[\bigg(\sum_{k=1}^{\infty}k\beta_0^{k-1} \big(\omega_0+ \alpha_0 X_{T-k}^2 \big)\bigg)^s\bigg] \\
\leq & \frac{1}{M^s}\EE \bigg[\sum_{k=1}^{\infty}k \beta_0^{s(k-1)} \big(\omega_0^s+ \alpha_0^s X_{T-k}^{2s} \big)\bigg] =  \frac{\omega_0^s+ \alpha_0^s \EE X_{t}^{2s}}{M^s(1-\beta_0^s)^2}
\end{split}
\end{align*}
such that $M>\Big(\frac{\omega_0^s+ \alpha_0^s \EE X_{t}^{2s}}{\epsilon(1-\beta_0^s)^2}\Big)^{1/s}$ establishes $\frac{\partial \psi(X_T,X_{T-1},\dots;\theta_0)}{\partial \beta}=O_p(1)$, which completes the verification of Assumption \ref{as:2.0}.c.\\

\noindent
Focusing on Assumption \ref{as:2.0}.d we notice that
\begin{align*}
\sup_{\theta \in \mathscr{V}(\theta_0)}\bigg|\frac{\partial^2 \psi(X_T,X_{T-1},\dots;\theta)}{\partial \omega^2}\bigg|=0,\\
\sup_{\theta \in \mathscr{V}(\theta_0)}\bigg|\frac{\partial^2 \psi(X_T,X_{T-1},\dots;\theta)}{\partial \omega \partial \alpha}\bigg|=0,\\
\sup_{\theta \in \mathscr{V}(\theta_0)}\bigg|\frac{\partial^2 \psi(X_T,X_{T-1},\dots;\theta)}{\partial \alpha^2}\bigg|=0
\end{align*}
and
\begin{align*}
\sup_{\theta \in \mathscr{V}(\theta_0)}\bigg|\frac{\partial^2 \psi(X_T,X_{T-1},\dots;\theta) }{\partial \omega \partial \beta}\bigg|= \frac{1}{(1-\beta_{\sup})^2}=O(1),
\end{align*}
where $\beta_{\sup}= \sup_{\theta \in \mathscr{V}(\theta_0)} \beta$.
To show  $\sup_{\theta \in \mathscr{V}(\theta_0)} \Big|\frac{\partial^2 \psi(X_T,X_{T-1},\dots;\theta)}{\partial \alpha \partial \beta}\Big|=O_{p}(1)$, we need to find an $M$ for every $\epsilon>0$  such that $\PP\Big[\sup_{\theta \in \mathscr{V}(\theta_0)}\Big|\frac{\partial^2 \psi(X_T,X_{T-1},\dots;\theta)}{\partial \alpha \partial \beta}\Big|\geq M\Big]<\epsilon$ holds. We find
\begin{align*}
&\PP\bigg[\sup_{\theta \in \mathscr{V}(\theta_0)}\bigg|\frac{\partial^2 \psi(X_T,X_{T-1},\dots;\theta) }{\partial \alpha \partial \beta}\bigg|\geq M\bigg]
\leq \frac{1}{M^s} \EE\Bigg[\bigg(\sum_{k=1}^\infty k\beta_{\sup}^{k-1}X_{T-k}^2\bigg)^s\Bigg]\\
\leq& \frac{1}{M^s} \EE\Bigg[\sum_{k=1}^\infty k^s\beta_{\sup}^{s(k-1)}X_{T-k}^{2s}\Bigg]\leq \frac{1}{M^s} \sum_{k=1}^\infty k\beta_{\sup}^{s(k-1)}\EE X_t^{2s}
= \frac{\EE X_t^{2s}}{M^s(1-\beta_{\sup}^s)^2} \:.
\end{align*}
Taking $M > \Big(\frac{\EE X_t^{2s}}{\epsilon(1-\beta_{\sup}^s)^2}\Big)^{1/s}$ leads to the desired result. Similarly, we have
%
%
\begin{align*}
\begin{split}
&\PP\bigg[\sup_{\theta \in \mathscr{V}(\theta_0)}\bigg|\frac{\partial^2 \psi(X_T,X_{T-1},\dots;\theta) }{\partial \beta^2 }\bigg|\geq M\bigg]\\
=&  \frac{1}{M^s}\EE \Bigg[\bigg(\sum_{k=2}^{\infty}k(k-1)\beta_{\sup}^{k-2} \big(\omega_{\sup}+\alpha_{\sup}  X_{T-k}^2\big)\bigg)^s\Bigg]\\
\leq&  \frac{1}{M^s}\EE \bigg[\sum_{k=2}^{\infty}k^s(k-1)^s\beta_{\sup}^{s(k-2)} \big(\omega_{\sup}^s+\alpha_{\sup}^s  X_{T-k}^{2s}\big)\bigg]\\
\leq&  \frac{1}{M^s} \sum_{k=2}^{\infty}k(k-1)\beta_{\sup}^{s(k-2)} \big(\omega_{\sup}^s+\alpha_{\sup}^s  \EE X_t^{2s}\big)=  \frac{2(\omega_{\sup}^s+\alpha_{\sup}^s \EE X_t^{2s})}{M^s(1-\beta_{\sup}^s)^3},
\end{split}
\end{align*}
where $\omega_{\sup}= \sup_{\theta \in \mathscr{V}(\theta_0)} \omega$ and $\alpha_{\sup}= \sup_{\theta \in \mathscr{V}(\theta_0)} \alpha$.
Taking $M>\Big(\frac{2(\omega_{\sup}^s+\alpha_{\sup}^s \EE X_t^{2s})}{\epsilon(1-\beta_{\sup}^s)^3}\Big)^{1/s}$ completes the verification of Assumption \ref{as:2.0}.d. \\

\noindent
Regarding Assumption \ref{as:2.0}.e we choose $\{c_t\}$ and $\{s_t\}$ to be sequences of zeros, i.e.\ $c_t=s_t=0$ for all $t \in \Z$, and note that
\begin{align*}
&\psi(X_T,X_{T-1},\dots;\theta) -\psi_{T+1}^s (\X_{t_1:T}^c; \theta) =   \sum_{k=T-t_1+1}^{\infty} \beta^k \alpha X_{T-k}^2.
\end{align*}
We have 
\begin{align}
\label{eq:7664489}
\begin{split}
&m_T\Big(\psi(X_T,X_{T-1},\dots;\theta_0) -\psi_{T+1}^s (\X_{t_1:T}^c; \theta_0)\Big) = \sqrt{T} \beta_0^{T-t_1}  \sum_{k=1}^{\infty} \beta_0^k \alpha_0 X_{t_1-k}^2.
\end{split}
\end{align}
Clearly, the sum is of order $O_p(1)$. Further, for any $t_1 \geq 1$ such that $(T-t_1) / l_T \rightarrow \infty$ we get $\sqrt{T} \beta_0^{T-t_1}\to 0$. Hence, \eqref{eq:7664489} is $o_p(1)$. Moreover, we obtain
\begin{align*}
&\bigg|\frac{\partial \psi_{T+1}^s (\X_{t_1:T}^c; \theta_0)}{\partial \omega} - \frac{\partial \psi (X_T, X_{T-1}, \ldots ;\theta_0)}{\partial \omega}\bigg|=  0
\end{align*}
and
\begin{align*}
&\bigg|\frac{\partial \psi_{T+1}^s (\X_{t_1:T}^c; \theta_0)}{\partial \alpha} - \frac{\partial \psi (X_T, X_{T-1}, \ldots ;\theta_0)}{\partial \alpha}\bigg| = \beta_0^{T-t_1} \sum_{k=1}^{\infty} \beta_0^k X_{t_1-k}^2
\end{align*}
being $o_p(1)$ since the sum is $O_p(1)$ and $\beta_0^{T-t_1}\to 0$. Similarly, we find
\begin{align*}
&\bigg|\frac{\partial \psi_{T+1}^s (\X_{t_1:T}^c; \theta_0)}{\partial \beta} - \frac{\partial \psi (X_T, X_{T-1}, \ldots ;\theta_0)}{\partial \beta}\bigg|=   \sum_{k=T-t_1+1}^{\infty} k \beta_0^{k-1} \alpha_0 X_{T-k}^2\\
= & (T-t_1) \beta_0^{T-t_1}  \sum_{k=1}^{\infty}  \beta_0^{k-1} \alpha_0 X_{t_1-k}^2+\beta_0^{T-t_1}  \sum_{k=1}^{\infty} k \beta_0^{k-1} \alpha_0 X_{t_1-k}^2
\end{align*}
being $o_p(1)$ and we conclude that
\begin{align*}
&\bigg|\bigg|\frac{\partial \psi_{T+1}^s (\X_{t_1:T}^c; \theta_0)}{\partial \theta} - \frac{\partial \psi (X_T, X_{T-1}, \ldots ;\theta_0)}{\partial \theta}\bigg|\bigg|=o_p(1).
\end{align*}
Further, we get
\begin{align*}
\sup_{\theta \in \mathscr{V}(\theta_0)}\bigg|\frac{\partial \psi_{T+1}^s (\X_{t_1:T}^c; \theta)}{\partial \omega \partial \theta'} - \frac{\partial \psi (X_T, X_{T-1}, \ldots ;\theta)}{\partial \omega \partial \theta'}\bigg|=&  (0,0,0),\\
\sup_{\theta \in \mathscr{V}(\theta_0)}\bigg|\frac{\partial \psi_{T+1}^s (\X_{t_1:T}^c; \theta)}{\partial \alpha^2} - \frac{\partial \psi (X_T, X_{T-1}, \ldots ;\theta)}{\partial \alpha^2}\bigg|=&0
\end{align*}
and
\begin{align*}
&\sup_{\theta \in \mathscr{V}(\theta_0)}\bigg|\frac{\partial \psi_{T+1}^s (\X_{t_1:T}^c; \theta)}{\partial \alpha \partial \beta} - \frac{\partial \psi (X_T, X_{T-1}, \ldots ;\theta)}{\partial \alpha \partial \beta}\bigg|= \sum_{k=T-t_1+1}^{\infty} k \beta_{\sup}^{k-1} X_{T-k}^2\\
= & (T-t_1) \beta_{\sup}^{T-t_1}  \sum_{k=1}^{\infty}  \beta_{\sup}^{k-1} X_{t_1-k}^2+\beta_{\sup}^{T-t_1}  \sum_{k=1}^{\infty} k \beta_{\sup}^{k-1} X_{t_1-k}^2
\end{align*}
is $o_p(1)$ by previous arguments noting that $\beta_{\sup} \in (0,1)$. Similarly, it can be shown that
\begin{align*}
&\sup_{\theta \in \mathscr{V}(\theta_0)}\bigg|\frac{\partial \psi_{T+1}^s (\X_{t_1:T}^c; \theta_0)}{\partial \beta \partial \beta'} - \frac{\partial \psi (X_T, X_{T-1}, \ldots ;\theta_0)}{\partial \beta \partial \beta'}\bigg|\\
=& \sup_{\theta \in \mathscr{V}(\theta_0)}\bigg|  \sum_{k=T-t_1+1}^{\infty} k(k-1) \beta^{k-2} \alpha X_{T-k}^2\bigg|\leq   \sum_{k=T-t_1+1}^{T-1} k(k-1) \beta_{\sup}^{k-2} \alpha_{\sup} X_{T-k}^2
\end{align*}
vanishes in probability to zero and we conclude that
\begin{align*}
\sup_{\theta \in \mathscr{V}(\theta_0)}\bigg|\bigg|\frac{\partial^2 \psi_{T+1}^s (\X_{t_1:T}^c; \theta)}{\partial \theta \partial \theta'} - \frac{\partial^2 \psi (X_T, X_{T-1}, \ldots ;\theta)}{\partial \theta \partial \theta'}\bigg|\bigg|=o_{p}(1).
\end{align*}

\subsubsection*{Assumption \ref{as:2.2}}

The condition in Assumption \ref{as:2.2}.a is satisfied for instance by $T_E(T)\sim T-\lfloor T^b \rfloor$ and  $T_P(T)\sim T-\lfloor T^a \rfloor$ with $0<a<b<1$.\\

\noindent
With regard to Assumption \ref{as:3.2}.4,  $\{X_t\}$ is a strictly stationary process such that Assumption \ref{as:2.2}.b is satisfied.\\

\noindent
The process $\{X_t\}$ is $\beta$-mixing with exponential decay  (\citeauthor{francq2011garchbook}, \citeyear{francq2011garchbook}, Thm.\ 3.4). As $\beta$-mixing implies $\alpha$-mixing (cf.\ \citeauthor{bradley2005basic}, \citeyear{bradley2005basic}), Assumption \ref{as:2.2}.c  is met with regard to remark 3 of BHS noting that $T_P(T)-T_E(T) \to \infty$.\\

\subsubsection*{Assumptions \ref{as:2.4} and \ref{as:2.5}}

Assumption \ref{as:2.4} is implied by Assumption \ref{as:2.5}, which, in turn, is verified by Theorem \ref{thm:3.2} and the consistent\footnote{A formal proof of consistency under Assumption \ref{as:3.2} is along the lines of the intermediary results (iv) and (vi) included in \citeauthor{francq2004maximum} (\citeyear{francq2004maximum}, Thm.\ 2.2).} estimator
\begin{align*}
\hat{\Upsilon}(\mathbf{X}_{1:T}) = \bigg(\frac{1}{T}\sum_{t=1}^T\frac{X_t^4}{\tilde{\sigma}_t^{\:4}\big(\hat{\theta}(\mathbf{X}_{1:T})\big)}-1\bigg)\:\bigg(\frac{1}{T}\sum_{t=1}^T\frac{1}{\tilde{\sigma}_t^{4}\big(\hat{\theta}(\mathbf{X}_{1:T})\big)}\frac{\partial \tilde{\sigma}_t^{\:2}\big(\hat{\theta}(\mathbf{X}_{1:T})\big)}{\partial \theta}\frac{\partial \tilde{\sigma}_t^{2}\big(\hat{\theta}(\mathbf{X}_{1:T})\big)}{\partial \theta'}\bigg)^{-1}.
\end{align*}

\subsubsection*{Assumptions within Corollary 1 of BHS}

To show $1/\hat{\upsilon}_T^{2IP}=O_p(1)$, recall that $\hat{\upsilon}_T^{2IP}= \upsilon_T^{2IP}+o_p(1)$ (see proof of Corollary 2 of BHS) and define $\kappa=eig_{\min}\Upsilon_0$, the   minimum eigenvalue of $\Upsilon_0$. Since $\Upsilon_0$ is positive definite, we have $\kappa>0$ such that $1/\hat{\upsilon}_T^{2IP}=O_p(1)$ is implied by
\begin{align*}
\hat{\upsilon}_T^{2IP}+o_p(1) = \upsilon_T^{2IP} \geq  \kappa \bigg|\bigg|\frac{\partial \psi_{T+1}}{\partial \theta}\bigg|\bigg|^2\geq \kappa \bigg|\frac{\partial \psi_{T+1}}{\partial \omega}\bigg|^2\geq \kappa\:.
\end{align*}
Similarly, we obtain $\hat{\upsilon}_T^{SPL}+o_p(1)\geq \kappa$ such that $1/\hat{\upsilon}_T^{SPL}=O_p(1)$.

\section{Conditional Mean in an ARMA$(1,1)$}
\label{sec:3.4}

\subsection{Model Description}
\label{sec:3.4.1}

The ARMA model was popularized by the classical book of \cite{box1971time}. It represents a stationary stochastic process in terms of an autoregressive and a moving-average part. The ARMA$(1,1)$ process with drift is given by
\begin{align}
\label{eq:3.4.1}
X_t-\omega_0 =  \alpha_0 \varepsilon_{t-1}+\beta_0 (X_{t-1}-\omega_0)+\varepsilon_t
\end{align}
for $t \in \Z$, where $\theta_0=(\omega_0,\alpha_0,\beta_0)'$ is a parameter vector in a parameter set $\Theta$ and $\{\varepsilon_t\}$ is a sequence of innovations. We denote by $\theta=(\omega,\alpha,\beta)'$ a generic  parameter vector and subsequently make the following assumptions:
\begin{assumption}{\textit{(ARMA$(1,1)$-Model)}}
\label{as:3.4}
\begin{enumerate}
\item[\ref{as:3.4}.1] \textit{(Compactness)} $\Theta$ is compact;
\item[\ref{as:3.4}.2] \textit{(Interior)} $\theta_0$ belongs to $\mathring{\Theta}$;
\item[\ref{as:3.4}.3] \textit{(Invertibility)}  $|\alpha|<1$  for all $\theta \in \Theta$;
\item[\ref{as:3.4}.4] \textit{(Causality)} $|\beta|< 1$ for all $\theta \in \Theta$;
\item[\ref{as:3.4}.5] \textit{(Roots)} $1-\beta z$ and $1+\alpha z$ have no common root, and $\alpha,\beta\neq 0$ for all $\theta \in \Theta$;
\item[\ref{as:3.4}.6] \textit{(Innovations)}
$\varepsilon_t$ are $\iid$ from an absolutely continuous distribution with respect to the Lebesgue measure on $\R$ satisfying  $\EE[\varepsilon_t]=0$,  $\EE[\varepsilon_t^2]=\sigma_\varepsilon^2<\infty$ and having a Lebesgue density strictly positive on $\R$;
\end{enumerate}
\end{assumption}
$\Theta$ is assumed to be compact in Assumption \ref{as:3.4}.1, which holds true, for instance, if it is of the form $\Theta=\big\{(\omega,\alpha,\beta)'\in \R^3: |\omega|\leq \delta^{-1},\delta\leq |\alpha|\leq 1-\delta \text{ and }\delta\leq |\beta|\leq 1-\delta\big\}$, where $\delta>0$ is a sufficiently small constant. Assumption \ref{as:3.4}.2 states that the true parameter vector lies in the interior of the parameter set and is necessary to obtain asymptotic normality of the parameter estimator. The invertibility and causality conditions are stated in \ref{as:3.4}.3 and \ref{as:3.4}.4. Assumption \ref{as:3.4}.5 ensures that the ARMA model is irreducible. Assumption \ref{as:3.4}.6 imposes further restrictions on the distribution of the innovation process. 
Next, we turn to the estimation of the model.

\subsection{Estimation}
\label{sec:3.4.2}

To estimate the model in equation \eqref{eq:3.4.1}, we consider a least squares estimator in the spirit of \cite{brockwell2013time}.\footnote{\cite{brockwell2013time} consider $\omega=0$ for simplicity. The extension to $\omega \neq 0$ is straight-forward.} Other estimators such as the QML estimator based on the Gaussian likelihood can alternatively be considered. Let $G_T(\alpha_0,\beta_0)$ be the correlation matrix of $(X_1,\dots,X_T)'$ with elements given by
\begin{align*}
\Cor(X_t,X_{t-k})=\frac{(\alpha_0+\beta_0)(1+\alpha_0\beta_0)}{1+2\alpha_0\beta_0+\alpha_0^2} \beta_0^{k-1}
\end{align*}
for $k\geq 1$. The (weighted) least squares estimator of $\theta_0$ is given by
\begin{align}
\label{eq:3.4.2}
\hat{\theta}(\mathbf{X}_{1:T})=&\arg\min_{\theta \in \Theta} (\mathbf{X}_{1:T}-\omega\iota_T)'
G_T^{-1}(\alpha,\beta)(\mathbf{X}_{1:T}-\omega \iota_T).
\end{align}
with $\iota_T=(1,\dots,1)'\in \R^T$. As the sample size grows large, the estimator approaches a normal distribution under regulatory conditions.

\begin{theorem}{\citep{brockwell2013time,bao2016asymptotic}} \label{thm:3.4}
Under Assumption \ref{as:3.4}
\begin{align}
\label{eq:3.4.3}
\sqrt{T}\big(\hat{\theta}(\mathbf{X}_{1:T})-\theta_0\big) \overset{d}{\to}N(0, \Upsilon_0)
\end{align}
with
\begin{align}
\label{eq:3.4.4}
\Upsilon_0=
\begin{pmatrix}
\frac{\sigma_\varepsilon^2(1-\alpha_0)^2}{(1-\beta_0)^2} &0 & 0\\
0 & \frac{(1-\alpha_0\beta_0)^2(1-\beta_0^2)}{(\alpha_0-\beta_0)^2} & \frac{(1-\alpha_0^2)(1-\alpha_0\beta_0)(1-\beta_0^2)}{(\alpha_0-\beta_0)^2} \\
0  & \frac{(1-\alpha_0^2)(1-\alpha_0\beta_0)(1-\beta_0^2)}{(\alpha_0-\beta_0)^2} & \frac{(1-\alpha_0\beta_0)^2(1-\alpha_0^2)}{(\alpha_0-\beta_0)^2}
\end{pmatrix}.
\end{align}
\end{theorem}
It is worth highlighting that $\Upsilon_0$ does not only depend on $\theta_0=(\omega_0,\alpha_0,\beta_0)'$, but also on the nuisance parameter $\sigma_{\varepsilon}^2$.

\subsection{Mapping}
\label{sec:3.4.3}

Having described the model and its estimation, we write the model in terms of the general framework. The conditional mean of  $X_{T+1}$ is equal to
\begin{align}
\label{eq:3.4.6}
\psi_{T+1} = \psi(X_T,X_{T-1},\dots;\theta_0)
= \omega_0 + \sum_{k=0}^\infty (-\alpha_0)^k (\alpha_0+\beta_0)(X_{T-k}-\omega_0).
\end{align}
To verify  Assumption \ref{as:2.0} requires the first and second derivatives of $\psi(X_T,X_{T-1},\dots;\theta)$ w.r.t.\ $\theta$. The first order derivatives are
\begin{align*}
\frac{\partial \psi(X_T,X_{T-1},\dots;\theta)}{\partial \omega}=& \frac{1-\beta}{1+\alpha},\\ 
\frac{\partial \psi(X_T,X_{T-1},\dots;\theta)}{\partial \beta}=& \sum_{k=0}^{\infty} (-\alpha)^{k} ( X_{T-k}-\omega),\\ 
\frac{\partial \psi(X_T,X_{T-1},\dots;\theta)}{\partial \alpha}=&  \sum_{k=0}^{\infty}(k+1)(-\alpha)^{k}\Big((X_{T-k}-\omega) - \beta (X_{T-k-1}-\omega)\Big),
\end{align*}
whereas the second order derivatives are given by
\begin{align*}
\frac{\partial^2 \psi(X_T,X_{T-1},\dots;\theta)}{\partial \omega^2}=&\: 0,\\
\frac{\partial^2 \psi(X_T,X_{T-1},\dots;\theta)}{\partial \beta^2}=&\: 0,\\
\frac{\partial^2 \psi(X_T,X_{T-1},\dots;\theta)}{\partial \omega \partial \beta}=&\:- \frac{1}{1+\alpha},\\
\frac{\partial^2 \psi(X_T,X_{T-1},\dots;\theta)}{\partial \omega \partial \alpha}=&\: -\frac{1-\beta}{(1+\alpha)^2},\\
\frac{\partial^2 \psi(X_T,X_{T-1},\dots;\theta)}{\partial \alpha \partial \beta}=&\sum_{k=0}^{\infty} (k+1)(-\alpha)^k (\omega- X_{T-k-1}),\\
\frac{\partial^2 \psi(X_T,X_{T-1},\dots;\theta)}{\partial \alpha^2}=& -\sum_{k=1}^{\infty}(k+1)k(-\alpha)^{k-1}\Big((X_{T-k}-\omega) - \beta (X_{T-k-1}-\omega)\Big).
\end{align*}

\subsection{Verification of Assumptions}
\label{sec:3.4.4}

Before turning to the verification of the high-level assumptions, note that $\EE|X_t|<\infty$ as the process $\{X_t\}$ is assumed to be causal and $\EE|\varepsilon_t|<\infty$.\footnote{As $\{X_t\}$ is causal, we can write it in the MA($\infty$) representation: $X_t-\omega_0=\sum_{j=0}^\infty\vartheta_j\varepsilon_{t-j}$ with $\sum_{j=0}^\infty|\vartheta_j|<\infty$ such that $\EE|X_t|\leq |\omega_0|+ \EE|\varepsilon_t|\sum_{j=0}^\infty|\vartheta_j|<\infty$.} 

\subsubsection*{Assumption \ref{as:2.0}}

For Assumption \ref{as:2.0}.a to be met, we consider the least squares estimator in equation  \eqref{eq:3.4.2}, whose asymptotic distribution is specified in Theorem \ref{thm:3.4}.\\

\noindent
 As the function $\psi_{T+1}(\dots;\theta)$ given in \eqref{eq:3.4.6} is continuous on $\Theta$ and twice differentiable on $\mathring{\Theta}$, Assumption \ref{as:2.0}.b is met.\\

\noindent
Regarding Assumption \ref{as:2.0}.c we note that
\begin{align*}
\frac{\partial \psi(X_T,X_{T-1},\dots;\theta_0)}{\partial \omega}= \frac{1-\beta_0}{1+\alpha_0}
\end{align*}
is trivially  $O(1)$. To show $\frac{\partial \psi(X_T,X_{T-1},\dots;\theta_0)}{\partial \beta}=O_p(1)$, we need to find a finite $M$ for every $\epsilon>0$ such that $\PP\big[\big|\frac{\partial \psi(X_T,X_{T-1},\dots;\theta_0)}{\partial \beta}\big|\geq M\big]<\epsilon$ for sufficiently large $T$. Employing the Markov inequality, we obtain
\begin{align*}
&\PP\bigg[\bigg|\frac{\partial \psi(X_T,X_{T-1},\dots;\theta_0)}{\partial \beta}\bigg|\geq M\bigg]\leq \frac{1}{M}\EE \Bigg[\bigg| \sum_{k=0}^{\infty} (-\alpha_0)^{k}  (X_{T-k}-\omega_0)\bigg|\Bigg]\\
\leq & \frac{1}{M}\sum_{k=0}^{\infty} |\alpha_0|^{k} \big( \EE|X_{t}|+|\omega_0|\big)= \frac{\EE|X_{t}|+|\omega_0|}{(1-|\alpha_0|)M}
\end{align*}
such that $M>\frac{\EE|X_{t}|+|\omega_0|}{(1-|\alpha_0|)\epsilon}$ gives the desired result. Similarly, we find
%
\begin{align*}
\begin{split}
&\PP\bigg[\bigg|\frac{\partial  \psi(X_T,X_{T-1},\dots;\theta_0)}{\partial \alpha}\bigg|\geq M\bigg]\\
\leq& \frac{1}{M}\EE \Bigg[\bigg|\sum_{k=0}^{\infty}(k+1)(-\alpha_0)^{k}\Big((X_{T-k}-\omega_0) - \beta_0 (X_{T-k-1}-\omega_0)\Big)\bigg|\Bigg] \\
\leq & \frac{1}{M}\EE \Bigg[\sum_{k=0}^{\infty}(k+1)|\alpha_0|^{k}\Big(|X_{T-k}|+|\omega_0| + |\beta_0|\: \big(|X_{T-k-1}|+|\omega_0|\big)\Big)\Bigg] \\
= & \frac{1}{M}\sum_{k=0}^{\infty}(k+1)|\alpha_0|^{k}\big(\EE|X_t|+|\omega_0|\big)(1 + |\beta_0|)=  \frac{(\EE|X_t|+|\omega_0|) (1+|\beta_0|)}{M(1-|\alpha_0|)^2}
\end{split}
\end{align*}
such that $M>\frac{(\EE|X_t|+|\omega_0|) (1+|\beta_0|)}{\epsilon(1-|\alpha_0|)^2}$ establishes $\frac{\partial \psi(X_T,X_{T-1},\dots;\theta_0)}{\partial \alpha}=O_p(1)$, which completes the verification of Assumption \ref{as:2.0}.c.\\

\noindent
Consider Assumption \ref{as:2.0}.d and note that 
\begin{align*}
\sup_{\theta \in \mathscr{V}(\theta_0)}\bigg|\frac{\partial^2 \psi(X_T,X_{T-1},\dots;\theta)}{\partial \omega^2}\bigg|=&0, \\
\sup_{\theta \in \mathscr{V}(\theta_0)}\bigg|\frac{\partial^2 \psi(X_T,X_{T-1},\dots;\theta)}{\partial \omega^2}\bigg|=&0,\\
\sup_{\theta \in \mathscr{V}(\theta_0)}\bigg|\frac{\partial^2 \psi(X_T,X_{T-1},\dots;\theta)}{\partial \omega \partial \beta}\bigg|=&\frac{1}{1-\alpha_{\sup}}=O(1)
\end{align*}
 and 
\begin{align*}
\sup_{\theta \in \mathscr{V}(\theta_0)}\bigg|\frac{\partial^2 \psi(X_T,X_{T-1},\dots;\theta)}{\partial \omega \partial \alpha}\bigg|\leq\frac{1+\beta_{\sup}}{1-\alpha_{\sup}},
\end{align*}
where $\alpha_{\sup}= \sup_{\theta \in \mathscr{V}(\theta_0)} |\alpha|$ as well as $\beta_{\sup}= \sup_{\theta \in \mathscr{V}(\theta_0)} |\beta|$.
%
%
%
%
To show that the term $\sup_{\theta \in \mathscr{V}(\theta_0)}\Big|\frac{\partial^2 \psi(X_T,X_{T-1},\dots;\theta)}{\partial \alpha \partial \beta}\Big|$ is $O_{p}(1)$, we need to find an $M$ for every $\epsilon>0$  such that $\PP\Big[\sup_{\theta \in \mathscr{V}(\theta_0)}\Big|\frac{\partial^2 \psi(X_T,X_{T-1},\dots;\theta)}{\partial \alpha \partial \beta}\Big|\geq M\Big]<\epsilon$ holds for sufficiently large $T$. We obtain
\begin{align*}
&\PP\bigg[\sup_{\theta \in \mathscr{V}(\theta_0)}\bigg|\frac{\partial^2 \psi(X_T,X_{T-1},\dots;\theta)}{\partial \alpha \partial \beta}\bigg|\geq M\bigg]\\
\leq& \frac{1}{M} \EE\Bigg[\sup_{\theta \in \mathscr{V}(\theta_0)}\bigg|\sum_{k=0}^{\infty} (k+1)(-\alpha)^{k} (\omega-X_{T-k-1})\bigg|\Bigg]\\
\leq& \frac{1}{M} \EE\Bigg[\sum_{k=0}^{\infty} (k+1)\alpha_{\sup}^{k}\big(|X_{T-k-1}|+\omega_{\sup}\big)\Bigg]\\
\leq & \frac{1}{M} \sum_{k=0}^{\infty} (k+1)\alpha_{\sup}^{k}\big(\EE|X_{t}|+\omega_{\sup}\big)
= \frac{\EE|X_{t}|+\omega_{\sup}}{M(1-\alpha_{\sup})^2},
\end{align*}
where $\omega_{\sup}= \sup_{\theta \in \mathscr{V}(\theta_0)} |\omega|$. Taking $M > \frac{\EE|X_{t}|+\omega_{\sup}}{\epsilon(1-\alpha_{\sup})^2}$ leads to the desired result. Similarly, we find
\begin{align*}
\begin{split}
&\PP\bigg[\sup_{\theta \in \mathscr{V}(\theta_0)}\bigg|\frac{\partial^2 \psi(X_T,X_{T-1},\dots;\theta)}{\partial \alpha^2 }\bigg|\geq M\bigg]\\
\leq&  \frac{1}{M}\EE \bigg[\sup_{\theta \in \mathscr{V}(\theta_0)} \bigg|\sum_{k=1}^{\infty}(k+1)k(-\alpha)^{k-1}\Big((X_{T-k}-\omega) - \beta (X_{T-k-1}-\omega)\Big)\bigg| \bigg]\\
\leq&  \frac{1}{M}\EE \bigg[ \sum_{k=1}^{\infty}(k+1) k \alpha_{\sup}^{k-1}\Big(|X_{T-k}|+\omega_{\sup} + \beta_{\sup} \big(|X_{T-k-1}|+\omega_{\sup}\big)\Big) \bigg]\\
\leq&  \frac{1}{M} \sum_{k=1}^{\infty}(k+1) k\alpha_{\sup}^{k-1}\big(\EE|X_{t}|+\omega_{\sup}\big)(1 + \beta_{\sup})
=  \frac{2(\EE|X_{t}|+\omega_{\sup})(1 + \beta_{\sup})}{M(1-\alpha_{\sup})^3}\:.
\end{split}
\end{align*}
Taking  $M> \frac{2(\EE|X_{t}|+\omega_{\sup})(1 + \beta_{\sup})}{\epsilon(1-\alpha_{\sup})^3}$ establishes that $\sup_{\theta \in \mathscr{V}(\theta_0)}\Big|\frac{\partial^2 \psi(X_T,X_{T-1},\dots;\theta)}{\partial \alpha^2 }\Big|=O_p(1)$.\\

\noindent
Regarding Assumption \ref{as:2.0}.e we choose $\{c_t\}$ and $\{s_t\}$ to be sequences of zeros, i.e.\ $c_t=s_t=0$ for all $t \in \Z$, and note that
\begin{align*}
&\psi(X_T,X_{T-1},\dots;\theta) -\psi_{T+1}^s (\X_{t_1:T}^c; \theta)
=  \sum_{k=T-t_1+1}^{\infty} (-\alpha_0)^k (\alpha_0+\beta_0)X_{T-k}.
\end{align*}
We have 
\begin{align}
\label{eq:787590805}
\begin{split}
&m_T\Big(\psi(X_T,X_{T-1},\dots;\theta_0) -\psi_{T+1}^s (\X_{t_1:T}^c; \theta_0)\Big)\\
=& \sqrt{T} (-\alpha_0)^{T-t_1}  
\sum_{k=1}^{\infty} (-\alpha_0)^k  (\alpha_0+\beta_0)X_{t_1-k}.
\end{split}
\end{align}
Clearly, the sum is of order $O_p(1)$ as $|\alpha_0|<1$ and $\{X_t\}$ is strictly stationary.  Further, for any $t_1 \geq 1$ such that $(T-t_1) / l_T \rightarrow \infty$ we get $\sqrt{T} (-\alpha_0)^{T-t_1}\to 0$. Hence, \eqref{eq:787590805} is $o_p(1)$. Moreover, we obtain
\begin{align*}
&\bigg|\frac{\partial \psi_{T+1}^s (\X_{t_1:T}^c; \theta_0)}{\partial \omega} - \frac{\partial \psi (X_T, X_{T-1}, \ldots ;\theta_0)}{\partial \omega}\bigg|=  0
\end{align*}
and
\begin{align*}
&\bigg|\frac{\partial \psi_{T+1}^s (\X_{t_1:T}^c; \theta_0)}{\partial \beta} - \frac{\partial \psi (X_T, X_{T-1}, \ldots ;\theta_0)}{\partial \beta}\bigg|\\
=&  \bigg|\sum_{k=T-t_1+1}^{\infty} (-\alpha_0)^k X_{T-k}\bigg|\leq   |\alpha_0|^{T-t_1}  
\sum_{k=1}^{\infty} |\alpha_0|^k |X_{t_1-k}|
\end{align*}
being $o_p(1)$ since the sum is $O_p(1)$ and $|\alpha_0|^{T-t_1}\to 0$. Similarly, we find
\begin{align*}
&\bigg|\frac{\partial \psi_{T+1}^s (\X_{t_1:T}^c; \theta_0)}{\partial \alpha} - \frac{\partial \psi (X_T, X_{T-1}, \ldots ;\theta_0)}{\partial \alpha}\bigg|\\
=&\bigg|    \sum_{k=T-t_1+1}^{\infty} \big((k+1)\alpha_0+k\beta_0\big)(-\alpha_0)^{k-1} X_{T-k}\bigg|\\
\leq &   \sum_{k=T-t_1+1}^{\infty} \big((k+1)|\alpha_0|+k|\beta_0|\big)|\alpha_0|^{k-1} |X_{T-k}|\\
\leq &  2 \big(|\alpha_0|+|\beta_0|\big) \sum_{k=T-t_1+1}^{\infty} k|\alpha_0|^{k-1} |X_{T-k}|\\
= &   (T-t_1)|\alpha_0|^{T-t_1}2 \big(|\alpha_0|+|\beta_0|\big)\sum_{k=1}^{\infty} |\alpha_0|^{k-1} |X_{t_1-k}|\\
&\qquad + |\alpha_0|^{T-t_1}2 \big(|\alpha_0|+|\beta_0|\big)\sum_{k=1}^{\infty} k|\alpha_0|^{k-1} |X_{t_1-k}|
\end{align*}
being $o_p(1)$ and we conclude that
\begin{align*}
&\bigg|\bigg|\frac{\partial \psi_{T+1}^s (\X_{t_1:T}^c; \theta_0)}{\partial \theta} - \frac{\partial \psi (X_T, X_{T-1}, \ldots ;\theta_0)}{\partial \theta}\bigg|\bigg|=o_p(1).
\end{align*}
Further, we get
\begin{align*}
\sup_{\theta \in \mathscr{V}(\theta_0)}\bigg|\frac{\partial \psi_{T+1}^s (\X_{t_1:T}^c; \theta)}{\partial \omega \partial \theta'} - \frac{\partial \psi (X_T, X_{T-1}, \ldots ;\theta)}{\partial \omega \partial \theta'}\bigg|&=  (0,0,0),\\
\sup_{\theta \in \mathscr{V}(\theta_0)}\bigg|\frac{\partial \psi_{T+1}^s (\X_{t_1:T}^c; \theta)}{\partial \beta^2} - \frac{\partial \psi (X_T, X_{T-1}, \ldots ;\theta)}{\partial \beta^2}\bigg|&=  0
\end{align*}
and
\begin{align*}
&\sup_{\theta \in \mathscr{V}(\theta_0)}\bigg|\frac{\partial \psi_{T+1}^s (\X_{t_1:T}^c; \theta)}{\partial \alpha \partial \beta} - \frac{\partial \psi (X_T, X_{T-1}, \ldots ;\theta)}{\partial \alpha \partial \beta}\bigg|\\
=& \sup_{\theta \in \mathscr{V}(\theta_0)}\bigg| \sum_{k=T-t_1}^{\infty} (k+1)(-\alpha)^k X_{T-k-1}\bigg|\leq  \sum_{k=T-t_1}^{\infty} (k+1)\alpha_{\sup}^k |X_{T-k-1}|\\
=& (T-t_1)\alpha_{\sup}^{T-t_1} \sum_{k=0}^{\infty}\alpha_{\sup}^k |X_{t_1-k-1}|+ \alpha_{\sup}^{T-t_1} \sum_{k=0}^{\infty} (k+1)\alpha_{\sup}^k |X_{t_1-k-1}|
\end{align*}
is $o_p(1)$ by previous arguments as $\alpha_{\sup} \in (0,1)$. Similarly, it can be shown that
\begin{align*}
&\sup_{\theta \in \mathscr{V}(\theta_0)}\bigg|\frac{\partial \psi_{T+1}^s (\X_{t_1:T}^c; \theta)}{\partial \alpha^2} - \frac{\partial \psi (X_T, X_{T-1}, \ldots ;\theta)}{\partial \alpha^2}\bigg|\\
=&\sup_{\theta \in \mathscr{V}(\theta_0)}\bigg|\sum_{k=T-t_1+1}^{\infty}(k+1)k(-\alpha)^{k-1} X_{T-k} - \beta \sum_{k=T-t_1}^{\infty}(k+1)k(-\alpha)^{k-1} X_{T-k-1}\bigg|\\
\leq &\sum_{k=T-t_1+1}^{\infty}(k+1)k\alpha_{\sup}^{k-1}|X_{T-k}| + \beta_{\sup} \sum_{k=T-t_1}^{\infty}(k+1)k\alpha_{\sup}^{k-1}|X_{T-k-1}|
\end{align*}
vanishes in probability to zero and we conclude that
\begin{align*}
\sup_{\theta \in \mathscr{V}(\theta_0)}\bigg|\bigg|\frac{\partial^2 \psi_{T+1}^s (\X_{t_1:T}^c; \theta)}{\partial \theta \partial \theta'} - \frac{\partial^2 \psi (X_T, X_{T-1}, \ldots ;\theta)}{\partial \theta \partial \theta'}\bigg|\bigg|=o_{p}(1).
\end{align*}

\subsubsection*{Assumption \ref{as:2.2}}

The condition in Assumption \ref{as:2.2}.a is satisfied for instance by $T_E(T)\sim T-\lfloor T^b \rfloor$ and  $T_P(T)\sim T-\lfloor T^a \rfloor$ with $0<a<b<1$.\\

\noindent
The process $\{X_t\}$ is strictly stationary since $|\beta_0|<1$ and $\EE\log^+|\varepsilon_t|\leq \EE|\varepsilon_t|<\infty$  (\citeauthor{bougerol1992stationarity}, \citeyear{bougerol1992stationarity}, Thm.\ 4.1).\\

\noindent
The process $\big\{(\varepsilon_t,X_t)\big\}$ is $\beta$-mixing with exponential decay  (\citeauthor{mokkadem1988mixing}, \citeyear{mokkadem1988mixing}, Thm.\ 1'). As $\beta$-mixing implies $\alpha$-mixing (cf.\ \citeauthor{bradley2005basic}, \citeyear{bradley2005basic}), Assumption \ref{as:2.2}.c  is met with regard to remark 3 of BHS noting that $T_P(T)-T_E(T) \to \infty$. For an alternative mixing result we refer to \citeauthor{davidson1994stochastic} (\citeyear{davidson1994stochastic}, Thm.\ 14.9).\\

\subsubsection*{Assumptions \ref{as:2.4} and \ref{as:2.5}}

Assumption \ref{as:2.4} is implied by Assumption \ref{as:2.5}, which, in turn,  is verified by Theorem \ref{thm:3.4} and the consistent estimator
\begin{align*}
\hat{\Upsilon}(\mathbf{X}_{1:T}) = \begin{pmatrix}
\hat{\Upsilon}_{11}(\mathbf{X}_{1:T}) &0 & 0\\
0 & \hat{\Upsilon}_{22}(\mathbf{X}_{1:T}) & \hat{\Upsilon}_{23}(\mathbf{X}_{1:T}) \\
0  & \hat{\Upsilon}_{23}(\mathbf{X}_{1:T}) & \hat{\Upsilon}_{33}(\mathbf{X}_{1:T})
\end{pmatrix},
\end{align*}
where 
\begin{align*}
\hat{\Upsilon}_{11}(\mathbf{X}_{1:T})=&\frac{\hat{\sigma}_\varepsilon^2(\mathbf{X}_{1:T})(1-\hat{\alpha}(\mathbf{X}_{1:T}))^2}{(1-\hat{\beta}(\mathbf{X}_{1:T}))^2}\\
\hat{\Upsilon}_{22}(\mathbf{X}_{1:T})=&\frac{(1-\hat{\alpha}(\mathbf{X}_{1:T})\hat{\beta}(\mathbf{X}_{1:T}))^2(1-\hat{\beta}(\mathbf{X}_{1:T})^2)}{(\hat{\alpha}(\mathbf{X}_{1:T})-\hat{\beta}(\mathbf{X}_{1:T}))^2}\\
\hat{\Upsilon}_{23}(\mathbf{X}_{1:T})=&\frac{(1-\hat{\alpha}(\mathbf{X}_{1:T})^2)(1-\hat{\alpha}(\mathbf{X}_{1:T})\hat{\beta}(\mathbf{X}_{1:T}))(1-\hat{\beta}(\mathbf{X}_{1:T})^2)}{(\hat{\alpha}(\mathbf{X}_{1:T})-\hat{\beta}(\mathbf{X}_{1:T}))^2}\\
\hat{\Upsilon}_{33}(\mathbf{X}_{1:T})=&\frac{(1-\hat{\alpha}(\mathbf{X}_{1:T})\hat{\beta}(\mathbf{X}_{1:T}))^2(1-\hat{\alpha}(\mathbf{X}_{1:T})^2)}{(\hat{\alpha}(\mathbf{X}_{1:T})-\hat{\beta}(\mathbf{X}_{1:T}))^2}
\end{align*}
and
\begin{align*}
\hat{\sigma}_\varepsilon^2(\mathbf{X}_{1:T})=&\frac{1}{T-3}\big(\mathbf{X}_{1:T}-\hat{\omega}(\mathbf{X}_{1:T})\iota_T\big)' G_T^{-1}\big(\hat{\alpha}(\mathbf{X}_{1:T}),\hat{\beta}(\mathbf{X}_{1:T})\big)\big(\mathbf{X}_{1:T}-\hat{\omega}(\mathbf{X}_{1:T})\iota_T\big).
\end{align*}

\subsubsection*{Assumptions within Corollary 1 of BHS}

To show $1/\hat{\upsilon}_T^{2IP}=O_p(1)$, recall that $\hat{\upsilon}_T^{2IP}= \upsilon_T^{2IP}+o_p(1)$ (see proof of corollary 2 of BHS) and define $\kappa=eig_{\min}\Upsilon_0$, the minimum eigenvalue of $\Upsilon_0$. Since $\Upsilon_0$ is positive definite, we have $\kappa>0$. Together with $|\beta_0|<1$ and $|\alpha_0|<1$
\begin{align*}
\hat{\upsilon}_n^{2IP}+o_p(1) =& \upsilon_n^{2IP}\geq \kappa \bigg|\bigg|\frac{\partial \psi_{T+1}^s(\X_{1:T};\theta_0)}{\partial \theta}\bigg|\bigg|^2\geq\kappa \bigg|\frac{\partial \psi_{T+1}^s(\X_{1:T};\theta_0)}{\partial \omega}\bigg|^2=\kappa \frac{(1-\beta_0)^2}{(1+\alpha_0)^2}> 0
\end{align*}
implies $1/\hat{\upsilon}_T^{2IP}=O_p(1)$. Analogously, we have $1/\hat{\upsilon}_T^{SPL}=O_p(1)$.

\section{Conditional Volatility in a T-GARCH(1,1)}
\label{sec:3.5}

\subsection{Model Description}
\label{sec:3.5.1}

The T-GARCH model was first introduced by \cite{zakoian1994threshold}. It accounts for the stylized fact that past positive and negative innovations appear not to have the same impact on current volatility, which is also known as leverage effect. The T-GARCH$(1,1)$ process $\{X_t\}$ is defined by 
\begin{align}
\label{eq:3.5.1}
\begin{split}
X_t =& \:\sigma_t \varepsilon_t\\
\sigma_t =&\: \omega_0 + \alpha_0^+ X_{t-1}^+ +\alpha_0^- X_{t-1}^- + \beta_0 \sigma_{t-1}
\end{split}
\end{align}
for all $t \in \Z$ using the notation $x^+=\max\{x,0\}$ and
$x^-=\max\{-x,0\}$. $\theta_0=(\omega_0,\alpha_0^+,\alpha_0^-,\beta_0)'$  are non-negative parameters in a parameter set $\Theta$ and $\{\varepsilon_t\}$ is a sequence of innovations. We denote by $\theta=(\omega,\alpha^+,\alpha^-,\beta)'$ a generic parameter vector and subsequently make the following assumptions:
%
\begin{assumption}{\textit{(T-GARCH(1,1)-Model)}}
\label{as:3.5}
\begin{enumerate}
\item[\ref{as:3.5}.1] \textit{(Compactness)} $\Theta$ is compact;
\item[\ref{as:3.5}.2] \textit{(Interior)} $\theta_0$ belongs to $\mathring{\Theta}$;
\item[\ref{as:3.5}.3] \textit{(Non-negativity)}  $\omega>0$, $\alpha^+\geq 0$, $\alpha^-\geq 0$  and  $ \beta \geq 0$ for all $\theta \in \Theta$;
\item[\ref{as:3.5}.4] \textit{(Strict Stationarity)} $\EE\big[\ln(\alpha_0^+ \varepsilon_t^+ + \alpha_0^- \varepsilon_t^- +\beta_0)\big]<1$ and $\beta<1$ for all $\theta \in \Theta$;
\item[\ref{as:3.5}.5] \textit{(Roots)} $1-\beta_0 z>0$ has no common root with $\alpha_0^+ z$ and $\alpha_0^- z$, and $\alpha_0^+ +\alpha_0^- \neq 0$;
\item[\ref{as:3.5}.6] \textit{(Innovations)}
$\varepsilon_t$ are $\iid$ from an absolutely continuous distribution with respect to the Lebesgue measure on $\R$ satisfying  $\EE[\varepsilon_t]=0$,  $\EE[\varepsilon_t^2]=1$ and $\EE[\varepsilon_t^4]<\infty$ and having a Lebesgue density strictly positive in a neighborhood of zero.
\end{enumerate}
\end{assumption}
$\Theta$ is assumed to be compact in Assumption \ref{as:3.5}.1, which holds true, for instance, if it is of the form $\Theta=[\delta,1/\delta]\times [0,1/\delta]^2\times [0,1-\delta]$, where $\delta \in (0,1)$ is a sufficiently small constant. Assumption \ref{as:3.5}.2 states that the true parameter vector lies in the interior of the parameter set and is necessary to obtain asymptotic normality of the parameter estimator.  The non-negativity constraints in \ref{as:3.5}.3 are standard ensuring the conditional standard deviation to be strictly positive. Assumption  \ref{as:3.5}.4 is necessary and sufficient for $\{X_t\}$ being strictly stationary (cf.\ \citeauthor{hamadeh2011asymptotic}, \citeyear{hamadeh2011asymptotic}). The root condition in \ref{as:3.5}.5 guarantees that the T-GARCH model is irreducible. Assumption \ref{as:3.5}.6 imposes further restrictions on the moments and density of the innovation process. 
Next, we turn to the estimation of the model in \eqref{eq:3.5.1}.

\subsection{Estimation}
\label{sec:3.5.2}

We consider the Gaussian QML estimator proposed by \cite{hamadeh2011asymptotic}. For a generic $\theta \in \Theta$ we set
\begin{align}
\label{eq:3.5.3}
 \sigma_{t+1}(\theta) = \sum_{k=0}^\infty \beta^k  \big(\omega +\alpha^+ X_{t-k}^+ + \alpha^- X_{t-k}^-\big)
\end{align}
and note that $\sigma_{t+1}=\sigma_{t+1}(\theta_0)$. Replacing the unknown presample observations by arbitrary values, say $s_t$, $t\leq 0$, we denote the modified version of \eqref{eq:3.5.3} by $\tilde{\sigma}_{t+1}^2(\theta)$.
%
%
Then the QML estimator of $\theta_0$ is defined as any measurable solution $\hat{\theta}(\mathbf{X}_{1:T})$ of
\begin{align}
\label{eq:6.2.4}
\hat{\theta}(\mathbf{X}_{1:T})=&\arg\max_{\theta \in \Theta} \tilde{L}_T(\theta;\mathbf{X}_{1:T})
\end{align}
with
\begin{align*}
\tilde{L}_T(\theta;\mathbf{X}_{1:T})=\prod_{t=1}^T\frac{1}{\sqrt{2\pi \tilde{\sigma}_t^2(\theta)}}\exp\left(-\frac{X_t^2}{2\tilde{\sigma}_t^2(\theta)}\right).
\end{align*}
%
%
Assumption \ref{as:3.5} implies that the  estimator follows asymptotically a normal distribution.
\begin{theorem}{\citep{hamadeh2011asymptotic}}
\label{thm:3.5}
Under Assumption \ref{as:3.5}
\begin{align}
\label{eq:3.5.5}
\sqrt{T}\big(\hat{\theta}(\mathbf{X}_{1:T})-\theta_0\big) \overset{d}{\to}N(0, \Upsilon_0)\:,
\end{align}
where $\Upsilon_0 = \frac{1}{4} \big(\EE[\varepsilon_t^4]-1\big)\:\EE\left[\frac{1}{\sigma_t^2(\theta_0)}\frac{\partial \sigma_t(\theta_0)}{\partial \theta}\frac{\partial \sigma_t(\theta_0)}{\partial \theta'}\right]^{-1}$ and $\sigma_t(\theta)$ is given in \eqref{eq:3.5.3}.
\end{theorem}
%

\subsection{Mapping}
\label{sec:3.5.3}

Having described the model and its estimation, we map the model into the general framework. The conditional volatility $\sigma_{T+1}$ is equal to
\begin{align}
\label{eq:3.5.7}
 \psi_{T+1}=\psi(X_T,X_{T-1},\dots;\theta_0) =& \sum_{k=0}^\infty \beta_0^k  \big(\omega_0 + \alpha_0^+ X_{T-k}^+ + \alpha_0^- X_{T-k}^-\big).
\end{align}
To verify Assumption \ref{as:2.0} the first and second derivatives of $\psi(X_T,X_{T-1},\dots;\theta)$  w.r.t.\ $\theta$ are needed. The first order derivatives are
\begin{align*}
\frac{\partial \psi(X_T,X_{T-1},\dots;\theta)}{\partial \omega}=& \frac{1}{1-\beta},\\
\frac{\partial \psi(X_T,X_{T-1},\dots;\theta)}{\partial \alpha^+}=& \sum_{k=0}^{\infty}\beta^k X_{T-k}^+,\\
\frac{\partial \psi(X_T,X_{T-1},\dots;\theta)}{\partial \alpha^-}=& \sum_{k=0}^{\infty}\beta^k X_{T-k}^-,\\
\frac{\partial \psi(X_T,X_{T-1},\dots;\theta)}{\partial \beta}=& \sum_{k=1}^{\infty}k\beta^{k-1} \big(\omega + \alpha^+ X_{T-k}^+ + \alpha^- X_{T-k}^-\big),
\end{align*}
whereas the second order derivatives are given by
\begin{align*}
\frac{\partial^2 \psi(X_T,X_{T-1},\dots;\theta)}{\partial \omega^2}=&\: 0 \\
\frac{\partial^2 \psi(X_T,X_{T-1},\dots;\theta)}{\partial \omega \partial \alpha^+}=&\: 0 \\
\frac{\partial^2 \psi(X_T,X_{T-1},\dots;\theta)}{\partial \omega \partial \alpha^-}=&\: 0 \\
\frac{\partial^2 \psi(X_T,X_{T-1},\dots;\theta) }{\partial \alpha^{+\:2}}=&\: 0 \\
\frac{\partial^2 \psi(X_T,X_{T-1},\dots;\theta)}{\partial \alpha^+ \partial \alpha^-}=&\: 0 \\
\frac{\partial^2 \psi(X_T,X_{T-1},\dots;\theta) }{\partial \alpha^{-\:2}}=&\: 0 \\
\frac{\partial^2 \psi(X_T,X_{T-1},\dots;\theta) }{\partial \omega \partial \beta}=&  \frac{1}{(1-\beta)^2}\\
%
\frac{\partial^2 \psi(X_T,X_{T-1},\dots;\theta) }{\partial \alpha^+ \partial \beta}=& \sum_{k=1}^{\infty}k \beta^{k-1} X_{T-k}^+\\
\frac{\partial^2 \psi(X_T,X_{T-1},\dots;\theta) }{\partial \alpha^- \partial \beta}=& \sum_{k=1}^{\infty}k \beta^{k-1} X_{T-k}^-\\
\frac{\partial^2 \psi(X_T,X_{T-1},\dots;\theta) }{\partial \beta^2}=& \sum_{k=2}^{\infty}k(k-1)\beta^{k-2} \big(\omega + \alpha^+ X_{T-k}^+ + \alpha^- X_{T-k}^-\big)
\end{align*}

\subsection{Verification of Assumptions}
\label{sec:3.5.4}

Before turning to the verification of the high-level assumptions, note that the strict stationarity condition implies the existence of fractional moments: there exists an $s\in (0,1)$ such that $E|X_t|^{s}<\infty$
(\citeauthor{hamadeh2011asymptotic}, \citeyear{hamadeh2011asymptotic}, Prop.\ A.1). 

\subsubsection*{Assumption \ref{as:2.0}}

For Assumption \ref{as:2.0}.a to be met, we consider the quasi-maximum likelihood estimator by \cite{hamadeh2011asymptotic}, whose asymptotic distribution is specified in Theorem \ref{thm:3.5}.\\

\noindent
As the function $\psi(\dots;\theta)$, given in \eqref{eq:3.5.7}, is continuous on $\Theta$ and twice differentiable on $\mathring{\Theta}$, Assumption \ref{as:2.0}.b is satisfied.\\

\noindent
Consider Assumption \ref{as:2.0}.c and note that
\begin{align*}
\frac{\partial \psi(X_T,X_{T-1},\dots;\theta_0)}{\partial \omega}=& \frac{1}{1-\beta_0}
\end{align*}
is trivally $O(1)$. For showing $\frac{\partial \psi(X_T,X_{T-1},\dots;\theta_0)}{\partial \alpha^+}=O_p(1)$, we need to find a finite $M$ for every $\epsilon>0$ such that $\PP\big[\big|\frac{\partial \psi(X_T,X_{T-1},\dots;\theta_0)}{\partial \alpha^+}\big|\geq M\big]<\epsilon$ for $T$ sufficiently large.  Markov's inequality implies
\begin{align*}
&\PP\bigg[\bigg|\frac{\partial \psi(X_T,X_{T-1},\dots;\theta_0)}{\partial \alpha^+}\bigg|\geq M\bigg]\leq \frac{1}{M^s}\EE \bigg[\bigg(\sum_{k=0}^\infty\beta_0^{k}X_{T-k}^+\bigg)^s\bigg]\\
\leq& \frac{1}{M^s}\EE \bigg[\bigg(\sum_{k=0}^\infty\beta_0^{k}|X_{T-k}|\bigg)^s\bigg]\leq\frac{1}{M^s}\sum_{k=0}^\infty\beta_0^{sk}\EE |X_{t}|^s
= \frac{\EE |X_t|^s}{(1-\beta_0^s)M^s}
\end{align*}
such that $M>\Big(\frac{\EE|X_t|^{s}}{(1-\beta_0^s)\epsilon}\Big)^{1/s}$ gives the desired result. The same $M$ serves to show $\frac{\partial \psi(X_T,X_{T-1},\dots;\theta_0)}{\partial \alpha^-}=O_p(1)$:
\begin{align*}
&\PP\bigg[\bigg|\frac{\partial \psi(X_T,X_{T-1},\dots;\theta_0)}{\partial \alpha^-}\bigg|\geq M\bigg]\leq \frac{1}{M^s}\EE \bigg[\bigg(\sum_{k=0}^\infty\beta_0^{k}X_{T-k}^-\bigg)^s\bigg]\\
\leq& \frac{1}{M^s}\EE \bigg[\bigg(\sum_{k=0}^\infty\beta_0^{k}|X_{T-k}|\bigg)^s\bigg]\leq \frac{\EE |X_t|^s}{(1-\beta_0^s)M^s}<\epsilon.
\end{align*}
Similarly, we get 
\begin{align*}
\begin{split}
&\PP\bigg[\bigg|\frac{\partial \psi(X_T,X_{T-1},\dots;\theta_0)}{\partial \beta}\bigg|\geq M\bigg] \leq  \frac{1}{M^s}\EE \bigg[\bigg(\sum_{k=1}^{\infty}k\beta_0^{k-1} \big(\omega_0 + \alpha_0^+ X_{T-k}^+ + \alpha_0^- X_{T-k}^-\big)\bigg)^s\bigg] \\
\leq & \frac{1}{M^s}\EE \bigg[\sum_{k=1}^{\infty}k \beta_0^{s(k-1)} \big(\omega_0^s+ (\alpha_0^+ + \alpha_0^-)^s |X_{T-k}|^s\big)\bigg] =  \frac{\omega_0^s+ (\alpha_0^+ + \alpha_0^-)^s \EE |X_{t}|^{s}}{M^s(1-\beta_0^s)^2}
\end{split}
\end{align*}
such that $M>\Big(\frac{\omega_0^s+ (\alpha_0^+ + \alpha_0^-)^s \EE |X_{t}|^s}{\epsilon(1-\beta_0^s)^2}\Big)^{1/s}$ establishes $\frac{\partial \psi(X_T,X_{T-1},\dots;\theta_0)}{\partial \beta}=O_p(1)$.\\

\noindent
Concerning Assumption \ref{as:2.0}.d we notice that
\begin{align*}
\sup_{\theta \in \mathscr{V}(\theta_0)}\bigg|\frac{\partial^2 \psi(X_T,X_{T-1},\dots;\theta)}{\partial \omega^2}\bigg|=&\: 0, \\
\sup_{\theta \in \mathscr{V}(\theta_0)}\bigg|\frac{\partial^2 \psi(X_T,X_{T-1},\dots;\theta)}{\partial \omega \partial \alpha^+}\bigg|=&\: 0, \\
\sup_{\theta \in \mathscr{V}(\theta_0)}\bigg|\frac{\partial^2 \psi(X_T,X_{T-1},\dots;\theta)}{\partial \omega \partial \alpha^-}\bigg|=&\: 0, \\
\sup_{\theta \in \mathscr{V}(\theta_0)}\bigg|\frac{\partial^2 \psi(X_T,X_{T-1},\dots;\theta) }{\partial \alpha^{+\:2}}\bigg|=&\: 0,\\
\sup_{\theta \in \mathscr{V}(\theta_0)}\bigg|\frac{\partial^2 \psi(X_T,X_{T-1},\dots;\theta)}{\partial \alpha^+ \partial \alpha^-}\bigg|=&\: 0, \\
\sup_{\theta \in \mathscr{V}(\theta_0)}\bigg|\frac{\partial^2 \psi(X_T,X_{T-1},\dots;\theta) }{\partial \alpha^{-\:2}}\bigg|=&\: 0 
\end{align*}
and
\begin{align*}
\sup_{\theta \in \mathscr{V}(\theta_0)}\bigg|\frac{\partial^2 \psi(X_T,X_{T-1},\dots;\theta) }{\partial \omega \partial \beta}\bigg|= \frac{1}{(1-\beta_{\sup})^2}=O(1),
\end{align*}
where $\beta_{\sup}= \sup_{\theta \in \mathscr{V}(\theta_0)} \beta$. To show  $\sup_{\theta \in \mathscr{V}(\theta_0)} \Big|\frac{\partial^2 \psi(X_T,X_{T-1},\dots;\theta)}{\partial \alpha^+ \partial \beta}\Big|=O_{p}(1)$, we need to find an $M$ for every $\epsilon>0$  such that $\PP\Big[\sup_{\theta \in \mathscr{V}(\theta_0)}\Big|\frac{\partial^2 \psi(X_T,X_{T-1},\dots;\theta)}{\partial \alpha^+ \partial \beta}\Big|\geq M\Big]<\epsilon$ holds. We obtain
\begin{align*}
&\PP\bigg[\sup_{\theta \in \mathscr{V}(\theta_0)}\bigg|\frac{\partial^2 \psi(X_T,X_{T-1},\dots;\theta) }{\partial \alpha^+ \partial \beta}\bigg|\geq M\bigg]
\leq \frac{1}{M^s} \EE\Bigg[\bigg(\sum_{k=1}^{\infty}k \beta_{\sup}^{k-1} X_{T-k}^+\bigg)^s\Bigg]\\
\leq& \frac{1}{M^s} \EE\Bigg[\sum_{k=1}^\infty k^s\beta_{\sup}^{s(k-1)}|X_{T-k}|^s\Bigg]\leq \frac{1}{M^s} \sum_{k=1}^\infty k\beta_{\sup}^{s(k-1)}\EE |X_{t}|^s
= \frac{\EE |X_{t}|^s}{M^s(1-\beta_{\sup}^s)^2}.
\end{align*}
Taking $M > \Big(\frac{\EE |X_t|^{s}}{\epsilon(1-\beta_{\sup}^s)^2}\Big)^{1/s}$ leads to the desired result. The same $M$ serves to prove that $\sup_{\theta \in \mathscr{V}(\theta_0)} \Big|\frac{\partial^2 \psi(X_T,X_{T-1},\dots;\theta)}{\partial \alpha^- \partial \beta}\Big|=O_{p}(1)$ since
\begin{align*}
&\PP\bigg[\sup_{\theta \in \mathscr{V}(\theta_0)}\bigg|\frac{\partial^2 \psi(X_T,X_{T-1},\dots;\theta) }{\partial \alpha^- \partial \beta}\bigg|\geq M\bigg]
\leq \frac{1}{M^s} \EE\Bigg[\bigg(\sum_{k=1}^{\infty}k \beta_{\sup}^{k-1} X_{T-k}^-\bigg)^s\Bigg]\\
\leq& \frac{1}{M^s} \EE\Bigg[\sum_{k=1}^\infty k^s\beta_{\sup}^{s(k-1)}|X_{T-k}|^s\Bigg]\leq \frac{\EE |X_{t}|^s}{M^s(1-\beta_{\sup}^s)^2}<\epsilon.
\end{align*}
Similarly, we have
\begin{align*}
\begin{split}
&\PP\bigg[\sup_{\theta \in \mathscr{V}(\theta_0)}\bigg|\frac{\partial^2 \psi(X_T,X_{T-1},\dots;\theta) }{\partial \beta^2 }\bigg|\geq M\bigg]\\
=&  \frac{1}{M^s}\EE \Bigg[\bigg(\sum_{k=2}^{\infty}k(k-1)\beta^{k-2} \big(\omega_{\sup}+\alpha_{\sup}^+  X_{T-k}^+ +\alpha_{\sup}^-  X_{T-k}^-\big)\bigg)^s\Bigg]\\
\leq&  \frac{1}{M^s}\EE \bigg[\sum_{k=2}^{\infty}k^s(k-1)^s\beta_{\sup}^{s(k-2)} \big(\omega_{\sup}^s+(\alpha_{\sup}^+ + \alpha_{\sup}^-)^s  |X_{T-k}|^{s}\big)\bigg]\\
\leq&  \frac{1}{M^s} \sum_{k=2}^{\infty}k(k-1)\beta_{\sup}^{s(k-2)} \big(\omega_{\sup}^s+(\alpha_{\sup}^+ + \alpha_{\sup}^-)^s   \EE |X_t|^{s}\big)=  \frac{2(\omega_{\sup}^s+(\alpha_{\sup}^+ + \alpha_{\sup}^-)^s   \EE |X_t|^{s})}{M^s(1-\beta_{\sup}^s)^3},
\end{split}
\end{align*}
where $\omega_{\sup}= \sup_{\theta \in \mathscr{V}(\theta_0)} \omega$, $\alpha_{\sup}^+= \sup_{\theta \in \mathscr{V}(\theta_0)} \alpha^+$ and $\alpha_{\sup}^-= \sup_{\theta \in \mathscr{V}(\theta_0)} \alpha^-$.
Taking $M>\Big(\frac{2(\omega_{\sup}^s+(\alpha_{\sup}^+ + \alpha_{\sup}^-)^s   \EE |X_t|^{s})}{\epsilon(1-\beta_{\sup}^s)^3}\Big)^{1/s}$ completes the verification of Assumption \ref{as:2.0}.d. \\

\noindent
Regarding Assumption \ref{as:2.0}.e we choose $\{c_t\}$ and $\{s_t\}$ to be sequences of zeros, i.e.\ $c_t=s_t=0$ for all $t \in \Z$, and note that
\begin{align*}
&\psi(X_T,X_{T-1},\dots;\theta) -\psi_{T+1}^s (\X_{t_1:T}^c; \theta)=  \sum_{k=T-t_1+1}^\infty \beta^k  \big( \alpha^+ X_{T-k}^+ + \alpha^- X_{T-k}^-\big).
\end{align*}
We have 
\begin{align}
\label{eq:75635789}
\begin{split}
&m_T\Big(\psi(X_T,X_{T-1},\dots;\theta_0) -\psi_{T+1}^s (\X_{t_1:T}^c; \theta_0)\Big)\\
&\qquad \qquad = \sqrt{T} \beta_0^{T-t_1} \sum_{k=1}^{\infty} \beta_0^k \big( \alpha_0^+ X_{t_1-k}^+ + \alpha_0^- X_{t_1-k}^-\big).
\end{split}
\end{align}
Clearly, the sum is of order $O_p(1)$. Further, for any $t_1 \geq 1$ such that $(T-t_1) / l_T \rightarrow \infty$ we get $\sqrt{T} \beta_0^{T-t_1}\to 0$. Hence, \eqref{eq:75635789} is $o_p(1)$. Moreover, we obtain
\begin{align*}
&\bigg|\frac{\partial \psi_{T+1}^s (\X_{t_1:T}^c; \theta_0)}{\partial \omega} - \frac{\partial \psi (X_T, X_{T-1}, \ldots ;\theta_0)}{\partial \omega}\bigg|=  0
\end{align*}
as well as 
\begin{align*}
&\bigg|\frac{\partial \psi_{T+1}^s (\X_{t_1:T}^c; \theta_0)}{\partial \alpha^+} - \frac{\partial \psi (X_T, X_{T-1}, \ldots ;\theta_0)}{\partial \alpha^+}\bigg| =  \sum_{k=T-t_1+1}^\infty \beta_0^k  X_{T-k}^+ \leq \beta_0^{T-t_1} \sum_{k=1}^{\infty} \beta_0^k X_{t_1-k}^+ 
\end{align*}
and
\begin{align*}
&\bigg|\frac{\partial \psi_{T+1}^s (\X_{t_1:T}^c; \theta_0)}{\partial \alpha^-} - \frac{\partial \psi (X_T, X_{T-1}, \ldots ;\theta_0)}{\partial \alpha^-}\bigg| =  \sum_{k=T-t_1+1}^\infty \beta_0^k  X_{T-k}^- \leq \beta_0^{T-t_1} \sum_{k=1}^{\infty} \beta_0^k X_{t_1-k}^- 
\end{align*}
being $o_p(1)$ since the sums are $O_p(1)$ and $\beta_0^{T-t_1}\to 0$. Similarly, we find
\begin{align*}
&\bigg|\frac{\partial \psi_{T+1}^s (\X_{t_1:T}^c; \theta_0)}{\partial \beta} - \frac{\partial \psi (X_T, X_{T-1}, \ldots ;\theta_0)}{\partial \beta}\bigg| =  \sum_{k=T-t_1+1}^\infty k \beta_0^{k-1}  \big( \alpha_0^+ X_{T-k}^+ + \alpha_0^- X_{T-k}^-\big)\\
= & (T-t_1) \beta_0^{T-t_1} \sum_{k=1}^{\infty}  \beta_0^{k-1}\big( \alpha_0^+ X_{t_1-k}^+ + \alpha_0^- X_{t_1-k}^-\big) +\beta_0^{T-t_1}  \sum_{k=1}^{\infty} k \beta_0^{k-1} \big( \alpha_0^+ X_{t_1-k}^+ + \alpha_0^- X_{t_1-k}^-\big)
\end{align*}
being $o_p(1)$ and we conclude that
\begin{align*}
&\bigg|\bigg|\frac{\partial \psi_{T+1}^s (\X_{t_1:T}^c; \theta_0)}{\partial \theta} - \frac{\partial \psi (X_T, X_{T-1}, \ldots ;\theta_0)}{\partial \theta}\bigg|\bigg|=o_p(1).
\end{align*}
Further, we get
\begin{align*}
\sup_{\theta \in \mathscr{V}(\theta_0)}\bigg|\frac{\partial \psi_{T+1}^s (\X_{t_1:T}^c; \theta)}{\partial \omega \partial \theta'} - \frac{\partial \psi (X_T, X_{T-1}, \ldots ;\theta)}{\partial \omega \partial \theta'}\bigg|&=  (0,0,0),\\
\sup_{\theta \in \mathscr{V}(\theta_0)}\bigg|\frac{\partial \psi_{T+1}^s (\X_{t_1:T}^c; \theta)}{\partial \alpha^{+\:2}} - \frac{\partial \psi (X_T, X_{T-1}, \ldots ;\theta)}{\partial \alpha^{+\:2}}\bigg|&=0,\\
\sup_{\theta \in \mathscr{V}(\theta_0)}\bigg|\frac{\partial \psi_{T+1}^s (\X_{t_1:T}^c; \theta)}{\partial \alpha^{-\:2}} - \frac{\partial \psi (X_T, X_{T-1}, \ldots ;\theta)}{\partial \alpha^{-\:2}}\bigg|&=0
\end{align*}
and
\begin{align*}
&\sup_{\theta \in \mathscr{V}(\theta_0)}\bigg|\frac{\partial \psi_{T+1}^s (\X_{t_1:T}^c; \theta)}{\partial \alpha^+ \partial \alpha^-} - \frac{\partial \psi (X_T, X_{T-1}, \ldots ;\theta)}{\partial \alpha^+ \partial \alpha^-}\bigg|=0.
\end{align*}
In addition, we find
\begin{align*}
&\sup_{\theta \in \mathscr{V}(\theta_0)}\bigg|\frac{\partial \psi_{T+1}^s (\X_{t_1:T}^c; \theta)}{\partial \alpha^+ \partial \beta} - \frac{\partial \psi (X_T, X_{T-1}, \ldots ;\theta)}{\partial \alpha^+ \partial \beta}\bigg|
 =  \sum_{k=T-t_1+1}^\infty k \beta_{\sup}^{k-1}  X_{T-k}^+ \\
= & (T-t_1) \beta_{\sup}^{T-t_1} \sum_{k=1}^{\infty}  \beta_{\sup}^{k-1} X_{t_1-k}^+ +\beta_{\sup}^{T-t_1}  \sum_{k=1}^{\infty} k \beta_{\sup}^{k-1}  X_{t_1-k}^+ 
\end{align*}
and
\begin{align*}
&\sup_{\theta \in \mathscr{V}(\theta_0)}\bigg|\frac{\partial \psi_{T+1}^s (\X_{t_1:T}^c; \theta)}{\partial \alpha^- \partial \beta} - \frac{\partial \psi (X_T, X_{T-1}, \ldots ;\theta)}{\partial \alpha^+ \partial \beta}\bigg|
 =  \sum_{k=T-t_1+1}^\infty k \beta_{\sup}^{k-1}  X_{T-k}^- \\
= & (T-t_1) \beta_{\sup}^{T-t_1} \sum_{k=1}^{\infty}  \beta_{\sup}^{k-1} X_{t_1-k}^- +\beta_{\sup}^{T-t_1}  \sum_{k=1}^{\infty} k \beta_{\sup}^{k-1}  X_{t_1-k}^- 
\end{align*}
being $o_p(1)$ by previous arguments noting that $\beta_{\sup} \in (0,1)$. Similarly, it can be shown that
\begin{align*}
&\sup_{\theta \in \mathscr{V}(\theta_0)}\bigg|\frac{\partial \psi_{T+1}^s (\X_{t_1:T}^c; \theta_0)}{\partial \beta \partial \beta'} - \frac{\partial \psi (X_T, X_{T-1}, \ldots ;\theta_0)}{\partial \beta \partial \beta'}\bigg|\\
\leq &  \sum_{k=T-t_1+1}^{\infty} k(k-1) \beta_{\sup}^{k-2} \big( \alpha_{\sup}^+ X_{T-k}^+ + \alpha_{\sup}^- X_{T-k}^-\big)
\end{align*}
vanishes in probability to zero and we conclude that
\begin{align*}
\sup_{\theta \in \mathscr{V}(\theta_0)}\bigg|\bigg|\frac{\partial^2 \psi_{T+1}^s (\X_{t_1:T}^c; \theta)}{\partial \theta \partial \theta'} - \frac{\partial^2 \psi (X_T, X_{T-1}, \ldots ;\theta)}{\partial \theta \partial \theta'}\bigg|\bigg|=o_{p}(1).
\end{align*}

\subsubsection*{Assumption \ref{as:2.2}}

The condition in Assumption \ref{as:2.2}.a is satisfied for instance by $T_E(T)\sim T-\lfloor T^b \rfloor$ and  $T_P(T)\sim T-\lfloor T^a \rfloor$ with $0<a<b<1$.\\

\noindent
With regard to Assumption \ref{as:3.5}.4,  $\{X_t\}$ is a strictly stationary process such that Assumption \ref{as:2.2}.b is satisfied.\\

\noindent
The process $\{X_t\}$ is $\beta$-mixing with exponential decay (\citeauthor{francq2006mixing}, \citeyear{francq2006mixing}, Thm.\ 3). As $\beta$-mixing implies $\alpha$-mixing (cf.\ \citeauthor{bradley2005basic}, \citeyear{bradley2005basic}), Assumption \ref{as:2.2}.c is met with regard to remark 3 of BHS noting that $T_P(T)-T_E(T) \to \infty$. For an alternative mixing result we refer to \cite{carrasco2002mixing}.\\

\subsubsection*{Assumptions \ref{as:2.4} and \ref{as:2.5}}

Assumption \ref{as:2.4} is implied by Assumption \ref{as:2.5}, which, in turn, is verified by Theorem \ref{thm:3.5} and the consistent estimator
\begin{align*}
\hat{\Upsilon}(\mathbf{X}_{1:T}) = \bigg(\frac{1}{T}\sum_{t=1}^T\frac{X_t^4}{\tilde{\sigma}_t^{\:4}\big(\hat{\theta}(\mathbf{X}_{1:T})\big)}-1\bigg)\:\bigg(\frac{1}{T}\sum_{t=1}^T\frac{1}{\tilde{\sigma}_t^{4}\big(\hat{\theta}(\mathbf{X}_{1:T})\big)}\frac{\partial \tilde{\sigma}_t^{\:2}\big(\hat{\theta}(\mathbf{X}_{1:T})\big)}{\partial \theta}\frac{\partial \tilde{\sigma}_t^{2}\big(\hat{\theta}(\mathbf{X}_{1:T})\big)}{\partial \theta'}\bigg)^{-1}.
\end{align*}

\subsubsection*{Assumptions within Corollary 1 of BHS}

The verification of $1/\hat{\upsilon}_T^{2IP}=O_p(1)$ and  $1/\hat{\upsilon}_T^{SPL}=O_p(1)$ is analogous to the GARCH($1,1$) case and hence omitted.

\section{Concluding Remarks}
\label{sec:3.6}

In this paper we establish the mapping of the conditional mean in an AR($1$) and ARMA($1,1$) model into the general setup. Further, the conditional variance and the conditional volatility in a GARCH($1,1$) and T-GARCH($1,1$) model, respectively, are shown to be encompassed in that framework. Further, the theoretical results of BHS are validated by verifying the corresponding assumptions for each model. Clearly, the list of nested models is non-exhaustive and can be extended. For instance one could study higher order models such as the ARMA($p,q$) or the GARCH($p,q$) model with $p,q \in \N$, which come at the cost of a more evolved analysis. Table \ref{tab:3.1} enlists four other GARCH-type extensions that are frequently encountered in the literature. 
\begin{table}
\centering
\begin{tabular}{cc}
\hline \hline
                 &    \\
Model            & Conditional variance $\sigma_t^2$ specification      \\
\hline
                 &    \\
E-GARCH($1,1$)   & $ \ln\sigma_t^2 = \omega + \alpha \frac{X_{t-1}}{\sigma_{t-1}}+\phi\Big(\Big|\frac{X_{t-1}}{\sigma_{t-1}}\Big|-\EE|\varepsilon_t|\Big)X_{t-1}^2 + \beta \ln \sigma_{t-1}^2$ \\
                 &   \\
N-GARCH($1,1$)   & $\sigma_t^2 = \omega+\alpha(X_{t-1}-\phi\sigma_{t-1})^2+\beta \sigma_{t-1}^2$   \\
                 &   \\
GJR-GARCH($1,1$) & $\sigma_t^2 = \omega + \alpha^+ X_{t-1}^2I_{\{X_{t-1}\geq 0\}} +\alpha^-,X_{t-1}^2 I_{\{X_{t-1}< 0\}} + \beta \sigma_{t-1}^2$                                               \\
                 &       \\
Q-GARCH($1,1$)   & $\sigma_t^2 = \omega + \alpha X_{t-1}^2 + \beta \sigma_{t-1}^2+\phi X_{t-1}$        \\
                 &       \\ \hline \hline
\end{tabular}
\caption{GARCH extensions. The respective process $\{X_t\}$ is generated by $X_t=\sigma_t \varepsilon_t$, where $\{\varepsilon_t\}$ is a sequence of innovations and $\sigma_t^2$ is the conditional variance at time $t$.}
\label{tab:3.1}
\end{table}
The family of quadratic GARCH (Q-GARCH) models has been proposed by \cite{sentana1995quadratic}. Its Q-GARCH($1,1$) member is very similar to the GARCH($1,1$) model and can be verified in a similar fashion replacing $\alpha X_{t-1}^2$ by $\alpha X_{t-1}^2+\phi X_t$. The GJR-GARCH($1,1$) model named after Glosten, Jagannathan and Runkle (\citeyear{glosten1993relation}) is a variant of the T-GARCH($1,1$), which corresponds to squaring the variables involved. It can be easily verified along the lines of Section \ref{sec:3.5}. The exponential GARCH (E-GARCH) model suggested by \cite{nelson1991conditional} and the non-linear GARCH (N-GARCH)  introduced by \cite{engle1993measuring} can also be embedded into the framework of BHS. For example, the conditional variance in an N-GARCH($1,1$) given by $\sigma_{T+1}^2 = \omega_0+\alpha_0(X_{T}-\phi_0\sigma_T)^2+\beta_0 \sigma_T^2 $,
%
where $\theta_0=(\omega_0, \alpha_0,\beta_0,\phi_0)'$ denotes the parameter vector. However, obtaining an explicit expression for the conditional variance in terms of $\theta_0$ and $\{X_t\}_{t\leq T}$ is complicated due to non-linearities in the recursive formula: e.g.\ $\sigma_{T+1}^2$ depends on $\sigma_T^2$ and $\sigma_T$ in the N-GARCH($1,1$).

There are few GARCH extensions such as the fractionally integrated (FI-GARCH) of \cite{baillie1996fractionally} or the fractionally integrated EGARCH (FIE-GARCH) of \cite{bollerslev1996modeling} that cannot be encompassed in the framework at hand.  The corresponding processes typically exhibit intermediate or long memory such that standard mixing results do not apply. Establishing the merging results on the basis of verifying Assumption \ref{as:2.2}.c  directly, instead via some mixing result, is an interesting question, which demands further investigation. 

Finally, we would like to emphasize that conditional risk measures such as conditional Value-at-Risk (VaR)  can be mapped into the general framework. For instance in the T-GARCH($1$,$1$) model of Section \ref{sec:3.5}, the conditional VaR of $X_{T+1}$ given $\{X_t\}_{t\leq T}$ at level $a \in (0,1)$ reduces to
\begin{align}
\label{eq:897780564}
VaR_a(X_{T+1}|X_T,X_{T-1},\dots) = -\xi_a \sum_{k=0}^\infty \beta_0^k  \big(\omega_0 + \alpha_0^+ X_{T-k}^+ + \alpha_0^- X_{T-k}^-\big)
\end{align}
with $\xi_a = \inf\big\{\tau \in \R :  \PP[\varepsilon_t \leq \tau]\geq a\big\}$; see \cite{francq2015risk} for details. Fixing $a$ and treating $\xi_a$ as additional parameter, \eqref{eq:897780564} is a function of $\{X_t\}_{t\leq T}$  and $\vartheta_0=(\omega_0,\alpha_0^+,\alpha_0^-,\beta_0, \xi_a)'$ and hence is nested in the setup of Section \ref{sec:3.0}. Similarly, the conditional Expected Shortfall (ES) of $X_{T+1}$ given $\{X_t\}_{t\leq T}$ at level $a \in (0,1)$
\begin{align}
ES_a(X_{T+1}|X_T,X_{T-1},\dots) = -\mu_a \sum_{k=0}^\infty \beta_0^k  \big(\omega_0 + \alpha_0^+ X_{T-k}^+ + \alpha_0^- X_{T-k}^-\big)
\end{align}
with $\mu_a = -\EE\big[\varepsilon_t|\varepsilon_t<\xi_a\big]$ can also be mapped into the general framework.

\singlespacing

\end{document}